\documentclass[twocolumn,aps,pra,amsmath,amssymb,superscriptaddress,showpacs]{revtex4-1}

\usepackage{graphicx}
\usepackage{graphics}
\usepackage{dcolumn}
\usepackage{bm}
\usepackage{color}
\usepackage{soul}
\usepackage{textcomp}
\usepackage{soul}
\usepackage{rotating}
\usepackage{setspace}
\usepackage[mathlines]{lineno}
\usepackage{enumerate}

\usepackage{microtype}
\usepackage[breaklinks=true,hyperindex=true,pdftitle={Quantum Driven Dissipative Parametric Oscillator in a Blackbody Radiation Field},
colorlinks=true,pagebackref=false,citecolor=blue,plainpages=false,pdfpagelabels,
linkcolor=blue,urlcolor=blue]{hyperref}

\def\bok#1#2#3{\left\langle#1\left|#2\right|#3 \right\rangle}
\def\sbok#1#2#3{\langle#1|#2|#3\rangle}

\def\iu{{\rm i}}

\begin{document}


\title{
Quantum Driven Dissipative Parametric Oscillator in a Blackbody Radiation Field
}

\author{Leonardo A. Pach\'on}
\affiliation{Grupo de F\'isica At\'omica y Molecular, Instituto de F\'{\i}sica,  Facultad de Ciencias Exactas y Naturales, 
Universidad de Antioquia UdeA; Calle 70 No. 52-21, Medell\'in, Colombia.}
\affiliation{Chemical Physics Theory Group, Department of Chemistry and
Center for Quantum Information and Quantum Control,
\\ University of Toronto, Toronto, Canada M5S 3H6}

\author{Paul Brumer}
\affiliation{Chemical Physics Theory Group, Department of Chemistry and
Center for Quantum Information and Quantum Control,
\\ University of Toronto, Toronto, Canada M5S 3H6}

\begin{abstract}
We consider the general open system problem of a charged quantum oscillator confined in a harmonic trap, whose frequency
can be arbitrarily modulated in time, that interacts with both an incoherent quantized (blackbody) 
radiation field and with an arbitrary coherent laser field.
We assume that the oscillator is initially in thermodynamic equilibrium with its environment,
a non-factorized initial density matrix of the system and the environment, and that at $t=0$ 
the modulation of the frequency, the coupling to the incoherent and the coherent radiation are 
switched on.
The subsequent dynamics, induced by the presence of the blackbody radiation, the
laser field and the frequency modulation, is studied in the framework of the influence functional approach.
This approach allows incorporating,  in  \textit{analytic closed formulae}, the non-Markovian 
character of the oscillator-environment
interaction at any temperature as well the non-Markovian character of the blackbody
radiation and its zero-point fluctuations.
Expressions for the time evolution of the covariance matrix elements of the
quantum fluctuations and the reduced density-operator are obtained. 
\end{abstract}

\date{\today}


\maketitle

\section{Introduction}
Since the seminal work of Magalinski{\u\i} \cite{Mag59}, Feynman and Vernon \cite{
FV63}, Ullersma \cite{Ull66,*Ull66b,*Ull66c,*Ull66d}, and  Caldeira-Leggett
\cite{CL83}, the theory of open quantum systems has been source of great interest
and object of continuous development, refinement and application (cf. Refs.~\citenum{
Wei08,GZ10,BP02,Muk99,MK11} and references therein).
This theory has, for example,  provided a solid conceptual framework  to explain 
fundamental phenomena such as the quantum-classical transition
\cite{Zur03}, violations of the second law of the thermodynamics \cite{EG&08}, the
survival of quantum features at high temperature \cite{GPZ10,PB11}, and has found 
applications in several fields in physics and chemistry \cite{Wei08,GZ10,BP02,Muk99,MK11}.

Despite the beauty and power of this theory, a study of the dynamics of a particular
system can be very cumbersome due to the sheer complexity of correctly incorporating
the various time, energy and coupling scales.
In order to circumvent this problem various approximations, such as the
weak-coupling-to-the-bath, Markovian, high temperature,  or
the initial factorizing condition is usually invoked \cite{GSI88,Wei08,GZ10,BP02,Muk99,MK11}.
However, the development of modern experimental techniques for preparing
and manipulating physical and chemical systems has reached the regime where
such approximations are questionable.
This fact has encouraged the development  of techniques for more consistently  treating
and analyzing open quantum systems  (cf. Ref.~\citenum{PB12c}).

Despite  the refinement in  technique, no approach  is completely
approximation-free \cite{Wei08,GZ10,BP02,Muk99,MK11,PB12c}, and therefore some 
physical features of the dissipative dynamics are often absent in these descriptions
\cite{Wei08,GZ10,BP02,PB12c,SB12}. 

The breakdown of these common approximations is expected to occur in the low
temperature regime \cite{HI05} and/or in the presence of initial correlations between
the system and the environment \cite{GSI88} and, in particular, in the case of driven 
non-equilibrium quantum systems \cite{ZH95,KK04,GPZ10,SN&11}.
The ubiquitous presence of this situation (an open quantum systems under the
presence of time dependent fields in, e.g., coherent control scenarios of chemical
systems \cite{SB11} and assorted  physical systems \cite{SN&11,PB13,PB12c}),
motivates a formal and detailed treatment of such dynamics.

In this paper, we derive formal \emph{exact} results for the non-Markovian
dynamics of a prototypical system, including the presence of initial correlations between
the system and the bath, and the possibility of arbitrary rapidly oscillating driving forces.
The method can be applied in the low temperature and strong coupling-to-the-bath regimes.
The particular system studied here is a charged quantum oscillator confined in a harmonic
trap that is  initially in thermodynamic equilibrium with its environment (non-factorized
initial conditions \cite{GSI88}).
For $t>0$, we start varying the frequency of the harmonic trap and couple the
oscillator, via the dipole, with an incoherent quantized blackbody
radiation field and with an arbitrary coherent laser field.
By means of the influence functional approach \cite{FV63,FH65,CL83,GSI88}, we
derive analytic closed expressions for the non-Markovian time evolution of the
covariance matrix elements of the quantum fluctuations valid at any temperature,
any system-environment coupling strength and incorporating the zero-point fluctuations
of the radiation.

This robust and general model allows us to address many different 
physical problems in generic open quantum systems.
The results derived here can be used to study, for example: 
\begin{enumerate}[i]
\item \emph{The incoherent excitation of open quantum systems}:
In particular, our results allow us to model situations when a molecular system 
such as retinal or a photosynthetic light-harvesting complex, equilibrated with it surroundings, is excited 
by either coherent (coherent laser pulses) or incoherent (sunlight or moonlight) light
sources, a subject of great significance in the chemical physics community \cite{JB91,BS11b,EC&07,HB11}.
One such application of this method is given in Ref.~\citenum{PB13}.

\item \emph{Environmentally-assisted one-photon phase control}:
That weak field 
one-photon phase control is not possible for certain isolated quantum systems
is a known result (cf. Refs.~\citenum{SB11,SAB10,PYB13} and references therein).
Recently, it was suggested that the coupling to the environment could allow, in principle,
for the coherent manipulation of quantum systems in such cases \cite{SAB10,PYB13}. 
However, it is still unclear what physical mechanisms are behind this process,
mainly because a formal study of this situation involves the presence of ultrafast field-induced 
modulations in open quantum systems at low temperature, a situation where non-Markovian 
processes cannot be disregarded and where approximations such as the rotating-wave 
fail \cite{KK04}.
Moreover, in this case of environmentally-assisted control,
the presence of initial correlations between the
system and the bath is vital. A treatment of this problem using the method developed
herein is to be provided in Ref. \citenum{PB13c}.
\item \emph{Optimal-control-based cooling of quantum nano-resonators by means of
parametric driving}.
On the basis of a numerical protocol, it was recently shown that nano-mechanical 
resonators can be cooled down by the delicate interplay of dissipative and driving
process \cite{SN&11}.
Being based on an optimal-control protocol, the possibility of parametrically driving the frequency 
of the resonator with arbitrary rapidly oscillating fields, as we consider here, is a key element 
in this cooling process.  
As in the previous case, this scenario is well beyond standard approximation schemes, but
can be immediately cast as a particular case of our general model, with the great advantage
of having an analytic formulae for the system dynamics.

\item \emph{The establishment of a quantum limit on non-Markovian time scales}.
In thermodynamical equilibrium, quantum features survive in the limit 
$\hbar \omega_0 / k_{\mathrm{B}}T > 1$,
where $\hbar \omega_0$ is a typical energy scale of the system and $T$ the temperature. 
According to Ref.~\citenum{GPZ10}, quantum features can persist for higher temperatures under
non-equilibrium situations.
However, results in Ref.~\citenum{GPZ10} are based on the Markovian approximation, so the
derivation a quantum limit consistent with the presence of non-Markovian effects is desirable.
This problem can be addressed using our general approach and this study is currently in progress.
\item \emph{Non-Markovian thermodynamics}.
For quantum systems, it is possible to have very fast control of heat and entropy due to
anomalies induced by the non-Markovian character of the relaxation \cite{EG&08}.
Our general results, can be immediately applied to study, e.g., heat transport between
two non-Markovian reservoirs at the quantum regime.
\end{enumerate}

This provides a sample list of problems that can be  
readily examined using the exact solution, derived below. It is the 
variety of challenging problems that can be addressed that motivates this paper,
the derivation
of an all-in-one versatile model that can treat a host of problems and which 
can be generalized to consider additional phenomena.

\section{Description of the System Hamiltonian}
The Hamiltonian of a quantum oscillator confined in a harmonic trap and in contact
with a thermal bath comprises three parts:
the Hamiltonian of the quantum oscillator itself, $\hat{H}_{\mathrm{S}}$;
the Hamiltonian of the thermal bath, $\hat{H}_{\mathrm{TB}}$, here described
according to the Ullersma-Caldeira-Leggett model \cite{Ull66,CL83}, as
a collection of harmonic modes;
and an interaction term $\hat{H}_{\mathrm{S-TB}}$ between the two systems.
For typical bilinear coupling, the three contributions are
\begin{align}
\label{equ:HS}
\hat{H}_{\mathrm{S}} &=
\frac{1}{2m} \hat{p}^2  + \frac{1}{2}m\omega(t)^2 \hat{q}^2  \\
\hat{H}_{\mathrm{TB}} &=
\sum_{j}^{\infty}\left[\frac{\hat{p}_j^2}{2m_j} + \frac{m_j \omega_j^2}{2} \hat{q}_j^ 2\right]
\\
\label{equ:HSTB}
\hat{H}_{\mathrm{S-TB}} &=
-\hat{q} \sum_{j}^{\infty}c_j \hat{q}_j
+ \hat{q}^2\sum_{j}^{\infty} \frac{c_j^2}{2 m_j \omega_j^2}
\end{align}
with $\hat{p}$ and $\hat{q}$ the canonically conjugate momentum and position of the 
oscillator (an analogous notation describes the bath modes),  $m$ the mass of the quantum oscillator and 
$\omega^2(t) = \omega_0^2 + \omega_{\mathrm{P}}^2(t)$ the parametrically 
modulated frequency. This frequency comprises two components: $\omega_0$,
a constant frequency and an arbitrary time-dependent-frequency $\omega_{\mathrm{P}}(t)$.
The magnitude of the interaction between the system and the bath is determined by the
coupling constants $c_j$.

In the presence of the blackbody radiation the Hamiltonian
$\hat{H}_0  = \hat{H}_{\mathrm{S}} +  \hat{H}_{\mathrm{S-TB}} +  \hat{H}_{\mathrm{TB}}$
needs to be augmented  to include the interaction with the field as well as the Hamiltonian
of the field modes,
\begin{equation}
\label{equ:HBB}
\hat{H}_{\mathrm{BB}} = \sum_{\mathbf{k},s}\hbar c k
   \left(\hat{a}_{\mathbf{k},s}^{\dagger} \hat{a}_{\mathbf{k},s}+\frac{1}{2}\right).
\end{equation}
Assuming that the charged oscillator interacts weakly with each mode in the field, we
can adopt a dipole-dipole type interaction, giving the overall Hamiltonian
\begin{linenomath}
\begin{equation}
\begin{split}
\label{equ:HamSysTBBB}
\hat{H} &= \frac{1}{2m}\left( \hat{p} - \frac{e}{c} \hat{A}\right)^2
   + \frac{1}{2}m\omega(t)^2 \hat{q}^2 
\\&
+ \hat{H}_{\mathrm{S-TB}} +  \hat{H}_{\mathrm{TB}}
+\hat{H}_{\mathrm{BB}},
\end{split}
\end{equation}
\end{linenomath}
where $e/c$ is the coupling constant to the radiation, $\hat{a}_{\mathbf{k},s}$ and
$\hat{a}_{\mathbf{k},s}^{\dagger}$ are the annihilation and creation operators of
the  field mode of momentum $\mathbf{k}$ and polarization $s$. The
vector potential is given by
\begin{linenomath}
\begin{align}
\label{equ:AVPBBR}
 \hat{A} &= \sum_{\mathbf{k},s} \left(\frac{h c}{kV}\right)^{\frac{1}{2}}
 f_k \vec{\mathbf{e}}_{\mathbf{k},s}
 \left(
    \hat{a}_{\mathbf{k},s}+\hat{a}_{\mathbf{k},s}^{\dagger}\right),
\end{align}
\end{linenomath}
where $\hat{\mathbf{e}}$ is the polarization unit vector, $V$ is the volume of the
auxiliary cavity containing the field modes and $f_k$ is the electron form-factor
(Fourier transform of the charge distribution) that incorporates the electron
structure \cite{FLO85,FLO88}. We have assumed, with no loss of generality,
that the form factor and polarization vector are real.
Note that by virtue of Eq.~(\ref{equ:HBB}), Eq.~(\ref{equ:HamSysTBBB}) already
contains zero-point or vacuum fluctuations.

Equation (\ref{equ:HamSysTBBB}) can be generalized to include  an additional term
$- \hat{q} E_{\mathrm{L}}(t)$, which allows for the possible manipulation of
the charged oscillator, via dipole coupling, by means of the electric field 
$E_{\mathrm{L}}(t)$ of a pulsed or continuous laser field.

Since Eq. (\ref{equ:HamSysTBBB}) includes the diamagnetic term
$\hat{A}^2$, it is not suitable for a path integration calculation,
which is  why it is usually omitted \cite{CK87,BC91}.
However, the contribution of this term is relevant for the derivation
of the partition function of the oscillator in the presence of the blackbody
radiation (cf. the discussion in \cite{FLO88b,CK88}).
In our case, this term can be introduced by means of the Power-Zienau's
transformation \cite{PZ59} (see also Ref.~\citenum{HG36} for the original version and 
Ref.~\citenum{YHJ81}
and references therein for a short historical review on this transformation),
$
\hat{T} = \exp\left\{\frac{\mathrm{i}}{\hbar}
\frac{e}{c}\hat{q} \cdot \hat{A}\right\},
$
which transforms
$\hat{p} \rightarrow \hat{p} + \frac{e}{c}\hat{A}$,
$\hat{q} \rightarrow \hat{q}$,
$\hat{p}_{\mathbf{k},s} \rightarrow \hat{p}_{\mathbf{k},s}
+ \frac{e}{c} m_{\mathbf{k}} \omega_{\mathbf{k}} \hat{q}$,
$\hat{q}_{\mathbf{k},s} \rightarrow \hat{q}_{\mathbf{k},s}$, where
we have defined \cite{FLO85,FLO88}
\begin{align}
\label{equ:AnhCreOpeField}
\hat{a}_{\mathbf{k},s} &=
(m_{\mathbf{k}} \omega_{\mathbf{k}} \hat{q}_{\mathbf{k},s}
+ \mathrm{i}\hat{p}_{\mathbf{k},s})/\sqrt{2m_{\mathbf{k}} \hbar \omega_{\mathbf{k}}},
\end{align}
with $m_{\mathbf{k}} = 4\pi e^2f_k^2/(\omega_{\mathbf{k}} V)$.
The corresponding total Hamiltonian reads
\begin{linenomath}
\begin{equation}
\begin{split}
\label{equ:HamSysTBBBb}
\hat{H} &= \frac{1}{2m} \hat{p}^2
 + \frac{1}{2}m\omega(t)^2 \hat{q}^2 - \hat{q} E_{\mathrm{L}}(t)
\\ & +  \sum_{j}\frac{\hat{p}_j^2}{2m_j}
+ \frac{m_j \omega_j^2}{2} \left(\hat{q}_j - \frac{c_j}{m_j \omega_j^2} \hat{q}\right)^2
\\ &
+ \sum_{\mathbf{k},s}
   \frac{1}{2 m_{\mathbf{k}}}\left(\hat{p}_{\mathbf{k},s} +
   m_{\mathbf{k}} \omega_{\mathbf{k}} \hat{q} \right)^2+
   \frac{1}{2}m_{\mathbf{k}}\omega_{\mathbf{k},s}^2 \hat{q}_{\mathbf{k},s}^2.
\end{split}
\end{equation}
\end{linenomath}
Here the oscillator is seen to be coupled to the momentum coordinate
$p_{\mathbf{k},s}$.

Recalling  that the electric field of the blackbody radiation is given by
\begin{linenomath}
\begin{equation}
\label{equ:DefQuaElecField}
\hat{E} = -\frac{\partial \hat{A}}{\partial t} =
 \textrm{i} \sum_{\mathbf{k},s} \left(\frac{h c^3}{V}\right)^{\frac{1}{2}}
 f_k \hat{\mathbf{e}}_{\mathbf{k},s}
 \left(
      \hat{a}_{\mathbf{k},s} - \hat{a}_{\mathbf{k},s}^{\dagger}\right),
\end{equation}
\end{linenomath}
and examining Eq.~(\ref{equ:AnhCreOpeField}), we see that in
Eq.~(\ref{equ:HamSysTBBBb}), the position of the oscillator is coupled
to the electric field of the blackbody radiation.
From an open-quantum-systems
perspective, this implies that any  statistical behavior induced by the blackbody
radiation (as seen elsewhere \cite{FLO85,FLO88,PB13})
is dictated by the stochastic fluctuations of the electric field.

\section{Initial Density Matrix}
After characterizing the Hamiltonian of the total system, the next step in the description 
of the dynamics is the determination of the initial state, taken to be the equilibrated state of S$+$TB.  
By denoting the coordinates of S$+$TB as
$\mathcal{Q} = \{q,\mathbf{Q}\}$,
the matrix elements of the initial density operator of the system S plus the environment TB
can be calculated as \cite{GSI88}
\begin{equation}
\label{equ:initialStateSTB}
\bok{\bar{\mathcal{Q}}'' }{\hat{\rho}_{\beta}}{\bar{\mathcal{Q}}'} =
Z_{\beta_{\mathrm{TB}}}^{-1}
\int \limits_{\bar{\mathcal{Q}}'}^{\bar{\mathcal{Q}}''}\mathcal{D} \bar{\mathcal{Q}}
\exp\left(- \frac{1}{\hbar}S^{\mathrm{E}}[\bar{\mathcal{Q}}]\right),
\end{equation}
where the integral is over all paths $\bar{\mathcal{Q}}(\tau)$,
$0 \leq \tau \leq \hbar\beta_{\mathrm{TB}}$ with
$\bar{\mathcal{Q}}(0) = \bar{\mathcal{Q}}'$
and $\bar{\mathcal{Q}}(\hbar \beta_{\mathrm{TB}}) = \bar{\mathcal{Q}}''$.
The bar stands for the trajectories $\mathcal{Q}(t)$ in imaginary time 
$t\rightarrow - \mathrm{i} \hbar \beta$, with $\beta_{\mathrm{TB}}=1/(k_{\mathrm{B}}T_{\mathrm{TB}})$.
In the imaginary-time path integral described by Eq.~(\ref{equ:initialStateSTB}),
$S^{\mathrm{E}}[\bar{\mathcal{Q}}]$ denotes the Euclidean action of
the system 
$S^{\mathrm{E}}[\bar{\mathcal{Q}}] = 
\int\limits_0^{\hbar \beta_{\mathrm{TB}}} \mathrm{d}\tau \mathcal{L}^{\mathrm{E}}(\bar{\mathcal{Q}},\dot{\bar{\mathcal{Q}}})
= S^{\mathrm{E}}_{\mathrm{S}}[\bar{q}] +
S^{\mathrm{E}}_{\mathrm{S-TB}}[\bar{q},\bar{\mathbf{Q}}]
+ S^{\mathrm{E}}_{\mathrm{TB}}[\bar{\mathbf{Q}}]$, 
obtained by introducing a global minus sign in the
potential energy \cite{GSI88,Ing02}. 
Since $\hat{\rho}_{\beta}$ denotes the density operator at $t=0$, we assume that
$\omega_{\mathrm{P}}(0)=0$ and $E_{\mathrm{L}}(0) = 0$ in Eq.~(\ref{equ:HamSysTBBBb}).
The matrix elements of the initial total density operator of interest are
\begin{equation}
\label{equ:initialState}
\begin{split}
\bok{\bar{q}_{\mathbf{k},s}'' \bar{\mathcal{Q}}'' }{\hat{\rho}(0)}{\bar{q}_{\mathbf{k},s}'\bar{\mathcal{Q}}'}&= 
\bok{\bar{q}_{\mathbf{k},s}''}{\hat{\rho}_{\beta_{\mathrm{BB}}}}{\bar{q}_{\mathbf{k},s}'}
\bok{\bar{\mathcal{Q}}'' }{\hat{\rho}_{\beta}}{\bar{\mathcal{Q}}'},
\end{split}
\end{equation}
where $\hat{\rho}_{\beta_{\mathrm{BB}}}$ denotes the equilibrium density operator of
the radiation only, at temperature $T_{\mathrm{BB}}$ and defined as in Eq.~(\ref{equ:initialStateSTB})
using the Euclidean action $S^{\mathrm{E}}_{\mathrm{BB}}[\bar{q}_{\mathbf{k},s}]$ of the
bare radiation, $\hat{\rho}_\beta$ is the thermal
density operator of system and bath introduced in Eq.~(\ref{equ:initialStateSTB}),
and the blackbody-radiation-mode-coordinates $q_{\mathbf{k},s}$ are defined 
in Eq.~(\ref{equ:AnhCreOpeField}).
When a system S is in contact with two thermal baths, what is usual in the literature 
(cf. Refs.~\citenum{PB13,PB13c}) is that the initial state of the total system is assumed to be 
factorized, in our case this is equivalent to take
$\hat{\rho}(0) = \hat{\rho}_{\beta_{\mathrm{BB}}} \otimes \hat{\rho}_{\beta_{\mathrm{TB}}}
\otimes \hat{\rho}_{\mathrm{S}}$, being $ \hat{\rho}_{\beta_{\mathrm{TB}}}$
and $ \hat{\rho}_{\mathrm{S}}$ the density operator of the thermal bath TB and
the system S, respectively. Here we deal with a more complex situation because
we take into account the initial correlations between the system S and the thermal bath 
TB.

In general, one would also like to study the system S prepared 
in a state other than  the
equilibrium state, e.g., in a coherent state or in a squeezed state. 
According to Refs.~\citenum{SG86,GSI88}, one could prepare a different
initial state by allowing the operators $\hat A_n, \hat A_n'$ acting only in the system
Hilbert space of S to generate an initial non-equilibrium density operator of system
and bath
$
\hat{\rho}_{\beta}^{\lambda} = \sum_n (\hat{A}_n \otimes \hat{1}_{\mathrm{TB}})
\hat{\rho}_\beta (\hat{1}_{\mathrm{TB}} \otimes \hat{A}_n')\,,
$
where $\hat{1}_{\mathrm{TB}}$ denotes the unit operator in the Hilbert space of the 
bath alone. 
In the position representation, the matrix elements of $\hat{\rho}_{\beta}^{\lambda}$
are given by
\begin{equation}
\label{equ:initialStateGen}
\begin{split}
\bok{\mathcal{Q}_+'}{\rho_{\beta}^{\lambda}}{\mathcal{Q}_-'}&=
\int \mathrm{d}\bar{\mathcal{Q}}''\mathrm{d}\bar{\mathcal{Q}}'
\lambda(q_{+}',{\bar q}'',q_{-}',{\bar q}')
\\&\times
\delta(\mathbf{Q}_+'-\bar{\mathbf{Q}}'') \delta(\mathbf{Q}_-'-\bar{\mathbf{Q}}')
\bok{\bar{\mathcal{Q}}'' }{\hat{\rho}_{\beta}}{\bar{\mathcal{Q}}'} ,
\end{split}
\end{equation}
where the propagating function
$
\lambda(q_{+}',\bar q'',q_{-}',\bar q')
=\sum_n \sbok{q_+'}{\hat{A}_n}{\bar{q}''} \sbok{\bar{q}'}{\hat{A}_n'}{q_{-}'}
$
characterizes the action of these operators. 
The delta functions  indicate that the imaginary-time paths for the bath degrees
of freedom are continuously connected to the real-time paths describing the time
evolution of the initial state \cite{SG86,GSI88}.
The thermal initial state in Eq.~(\ref{equ:initialStateSTB}) can be recovered
by setting the auxiliary operators $\hat{A}_n$ and $\hat{A}_n'$ to $\hat{1}_{\mathrm{S}}$, 
which yields $\lambda(q_{+}',{\bar q}'',q_{-}',{\bar q}') = \delta(q_{+}'-{\bar q}'')
\delta(q_{-}'-{\bar q}')$.
As distinct from $\bar{\mathcal{Q}}'$ and $\bar{\mathcal{Q}}''$, which are the endpoints
of a single imaginary-time-trajectory, $\mathcal{Q}_{+}'$ and $\mathcal{Q}_{-}'$ denote 
the initial condition for two real-time-trajectories ``$+$'' and ``$-$'' 
(see below). For
the case of unitary evolution these can be identified as the forward and backward trajectories
associate to the unitary time-evolution-operator and its adjoint, respectively.
Additional details about different initial preparations can be found in 
Refs.~\citenum{SG86,GSI88}.

\section{Dynamics of the System}
In this section we solve for the time evolution of the initial density matrix
[Eq.~(\ref{equ:initialState})] under the action of the Hamiltonian
[Eq.~(\ref{equ:HamSysTBBBb})], using the Feynman and Vernon
influence functional approach \cite{FV63,FH65}. For this problem we require  a
mixture of the influence functional for factorizing initial conditions
\cite{FV63,FH65,CL83,ZH95}  in order to consider the effect of the radiation, and for
non-factorizing initial conditions \cite{SG86,GSI88} in order to correctly characterize
the equilibrium state between the oscillator and the thermal bath at $t=0$.

It is worth mentioning that the standard path integral calculations are performed for position-position 
couplings \cite{FV63,FH65,FV63,FH65,CL83,ZH95,SG86,GSI88,Wei08} [cf. the term
$\hat{q}\, \hat{q}_j$ in Eq.~(\ref{equ:HamSysTBBBb})], and for momentum-momentum
coupling \cite{CL83}.
The oscillator-radiation coupling is of position-momentum type,
$\hat{q}\,\hat{p}_{\mathbf{k},s}$ [cf. Eq.~(\ref{equ:HamSysTBBBb})].
By means of a set of unitary transformations \cite{FLO88}, one could invert the
role of the momentum and position operators of the field [see
Eq.~\ref{equ:AnhCreOpeField}], with the caveat that this generates an additional
term proportional to the initial position of the oscillator (see below).
However, since the system described in Eq.~(\ref{equ:HamSysTBBBb}) is still linear,
the path integral calculation can  also be carried out analytically for the
position-momentum coupling.  This is the approach  followed below.

\subsection{Derivation of the propagating function and the influence functional}
The time evolution of the system S is described by the reduced density operator 
$\hat{\rho}_{\mathrm{S}}(t) = \mathrm{tr}_{\mathrm{TB},\mathrm{BB}}\hat{\rho(t)}$. 
Following Ref.~\citenum{GSI88}, we obtain that the matrix elements of $\hat{\rho}_{\mathrm{S}}(t)$
are given by
\begin{equation}
\label{equ:ForTimeEvo}
\begin{split}
\bok{q_+''}{\hat{\rho}_{\mathrm{S}}(t)}{q_-''} &=
\int  \mathrm{d}q_+' \mathrm{d}q_-' \mathrm{d}\bar{q}'' \mathrm{d}\bar{q}'
\lambda(q_{+}',{\bar q}'',q_{-}',{\bar q}')
\\&\times
J(q_+'',q_-'',t;q_+',q_-',0;\bar{q}'',\bar{q}') ,
\end{split}
\end{equation}
where $J(q_+'',q_-'',t;q_+',q_-',0;\bar{q}'',\bar{q}')$ is the propagating function of the
system density matrix which can be expressed in terms of the functional
phase $\Sigma[q_+,q_-,\bar{q}]$ by means of the three-fold path integral
expression
\begin{equation}
\label{equ:GenPropFunct}
\begin{split}
J(q_+''&,q_-'',t;q_+',q_-',0;  \bar{q}'',\bar{q}')  =
\\&
\frac{1}{Z} \int\limits_{q_+'}^{q_+''} \mathcal{D}q_+
\int\limits_{q_-'}^{q_-''} \mathcal{D}q_-
\int\limits_{\bar{q}'}^{\bar{q}''}  \mathcal{D}\bar{q}
\exp\left(\frac{\mathrm{i}}{\hbar} \Sigma[q_+,q_-,\bar{q}]\right),
\end{split}
\end{equation}
where $Z$ normalizes $J(q_+'',q_-'',t;q_+',q_-',0;  \bar{q}'',\bar{q}')$ to
$\delta(q_+''-q_+')\delta(q_-''-q_-') \sbok{\bar{q}''}{\hat{\rho}_{\beta,\mathrm{S}}}{\bar{q}'}$ 
at $t=0$, being $\hat{\rho}_{\beta,\mathrm{S}} = \mathrm{tr}_{\mathrm{TB}}(\hat{\rho}_{\beta})$.
The real time path-integrals over $q_+$ and $q_-$ are carried out subject
to the endpoints $q_+(0) = q_+'$, $q_+(t) = q_+''$, $q_-(0) = q_-'$ and
$q_-(t) = q_-''$, while the imaginary time path integral are over
$\bar{q}(0)=\bar{q}'$ and $\bar{q}(\hbar\beta_{\mathrm{TB}})=\bar{q}''$.
Recall that the imaginary time path-integral allows for the calculation of the equilibrated 
density operator of S$+$TB and the influence of their initial correlations in the 
subsequence time evolution. 

After tracing over the degree of freedom of TB and BB, and after defining
$q_{+} = (r + x)/2$ and $q_{-} = r - x$, we have that the functional phase
$\Sigma [x,r,\bar{q}]$ is given by
\begin{widetext}
\begin{equation}
\label{equ:GenSigma}
\begin{split}
&\Sigma [x,r,\bar{q}] =
\\
& \mathrm{i} \int\limits_0^{\hbar \beta_{\mathrm{TB}}} \mathrm{d}\tau
\left[
\frac{m}{2} \dot{\bar{q}}^2 + \frac{1}{2}m\omega_0^2 \bar{q}^2 + \frac{1}{2}
\int\limits_0^{\hbar \beta_{\mathrm{TB}}}\mathrm{d}\sigma
k_{\mathrm{TB}}(\tau - \sigma)\bar{q}(\tau) \bar{q}(\sigma)
\right]
+
\int\limits_0^{\hbar \beta_{\mathrm{TB}}}\mathrm{d}\tau \int\limits_0^{t}\mathrm{d} s
K_{\mathrm{TB}}^*(s - \mathrm{i} \tau)\bar{q}(\tau) x(s)
\\
& + \int\limits_0^{t} \mathrm{d}s \left\{m \dot{x}(s) \dot{r}(s) - m \omega_0^2 r(s) x(s)
- m \omega_{\mathrm{P}}(t)^2 r(s) x(s) + E_{\mathrm{L}}(t) x(s)
- r'\eta_{\mathrm{TB}}(s)x(s)\right\}
\\
\phantom{\Sigma [x,r,\bar{q}] =}
&- \int\limits_0^{t}\mathrm{d}s
\left\{
\int\limits_0^{t}\mathrm{d} u
\left[ \eta_{\mathrm{TB}}(s) + \eta_{\mathrm{BB}}(s)\right] x(s) \dot{r}(u)
- \frac{\mathrm{i}}{2} \int\limits_0^{t}\mathrm{d} u
\left[ K^{\mathrm{re}}_{\mathrm{TB}}(s-u) + K_{\mathrm{BB}}(s-u)\right] x(s) x(u)
\right\}.
\end{split}
\end{equation}
\end{widetext}
Note that $r' = q_+' + \frac{1}{2}q_-'$, $r'' = q_+'' + \frac{1}{2}q_-''$ and 
analogously  for $x''$ and $x'$.
The various kernels entering into Eq. (\ref{equ:GenSigma}) are defined
in the next section.

The first term in the first line of Eq. (\ref{equ:GenSigma}) accounts  for the
equilibrium density operator of S in the presence of the thermal bath TB while
the second term containing $\bar{q}(\tau)x(s)$ is responsible of the the effect 
of initial correlations between the environment and the system on the 
subsequent time evolution.
The first three terms in the second line are responsible of the evolution under
 the parametric harmonic potential, while the fourth term is
responsible for the evolution induced by the laser field $E_{\mathrm{L}}$.
As previously noted, we take $\omega_{\mathrm{P}}(0) =0$ and $E_{\mathrm{L}}(0) =0$.
The last term in the second line of Eq.~(\ref{equ:GenSigma}) arises  from
the incoherent excitation induced by the position-position coupling to the
thermal bath TB.
Since the coupling to the blackbody radiation field BB is of a different nature,
position-momentum coupling, this transient term proportional to the initial
position $r'$ is not present.  However, if one changes the role of position
and momentum, as discussed above, this transient term enters implicitly.
The terms in the third line constitute the exponent of the influence functional
of the Feynman-Vernon theory under the action of the thermal
bath TB and  blackbody radiation BB.

The additional time integration in the last line of Eq.~(\ref{equ:GenSigma}) over $u$
accounts for the non-local time (non-Markovian) evolution of the density operator.
Although the temporal non-locality is determined by the various kernels in a
cumbersome way, we can identify two kinds of non-Markovian contributions:
 one from the dissipative part and determined by the non-local
character of $\eta_{\mathrm{TB}}$ and $\eta_{\mathrm{BB}}$, and
 a second  determined by the thermal fluctuations  described (see below)
by the kernels $K^{\mathrm{re}}_{\mathrm{TB}}$ and $K_{\mathrm{BB}}$.
The presence of the latter is not determined by the
presence of the former, i.e., in the limit of local dissipative Ohmic kernels,
$\eta_{\mathrm{TB}}(s) \sim \delta(s)$, the non-local character of the thermal
fluctuations is still present; it only vanishes in the high temperature regime
\cite{Wei08,Pac10}.
\subsection{Kernels in the functional action}
The quantities introduced in the effective action \( \Sigma [x,r,\bar{q}_x]\)
are defined in terms of the bath spectral density $J_{\mathrm{TB}}$
and the blackbody-radiation spectral density $J_{\mathrm{BB}}$.
These spectral densities are determined  \cite{CL83,GSI88,FLO85,FLO88} from the parameters of the bath
modes and the coupling constants by means of
\begin{align}
\label{equ:DefJwTB}
J_{\mathrm{TB}}(\omega)&= \frac{\pi}{2} \sum_{j=1}^{\infty} \frac{c_j^2}{m_j\omega_j}
\delta(\omega - \omega_j),
\\
\label{equ:DefJwBB}
J_{\mathrm{BB}}(\omega)&= \frac{\pi}{2} \sum_{\mathbf{k},s}^{\infty}
m_{\mathbf{k}}\omega_{\mathbf{k}}^3
\delta(\omega - \omega_{\mathbf{k}}).
\end{align}
Assuming that the thermal bath is dense in the frequency of the modes
\cite{CL83}, it is customary to describe the spectral density in Eq.~(\ref{equ:DefJwTB})
by, e.g., assuming the Ohmic model
\begin{equation}
\label{equ:JwTB}
J_{\mathrm{TB}}(\omega) = m \gamma_{\mathrm{TB}} \omega
\Omega_{\mathrm{TB}}^2/(\Omega_{\mathrm{TB}}^2 + \omega^2 ),
\end{equation}
where $\gamma_{\mathrm{TB}}$ is the coupling constant to the bath TB and
$\Omega_\mathrm{TB}$ is a cutoff parameter related to the inverse of
the bath memory time.
In contrast with the thermal bath case, no assumption on the functional
form of the spectral density of the blackbody is needed in the continuous
limit \cite{FLO85,FLO88,FLO87,BC91}.

The transversality condition implies  that only two of the three components
of $\mathbf{k}$ contribute to the coupling \cite{FLO88}, giving a global
factor of two-thirds for the spectral density in Eq.~(\ref{equ:DefJwBB}).
In the continuous limit,
$\sum_{\mathbf{k}}\rightarrow \frac{V}{(2\pi)^3}\int\mathrm{d}\mathbf{k}$,
the spectral density for the blackbody radiation is
\begin{equation}
\label{equ:JwBB}
J_{\mathrm{BB}}(\omega) = M \tau_{\mathrm{BB}} \, \omega^3 \Omega_{\mathrm{BB}}^2/
\left(\Omega_{\mathrm{BB}}^2 + \omega^2\right),
\end{equation}
where $M = m + M \tau_{\mathrm{BB}} \Omega_{\mathrm{BB}}$ is the renormalized mass,
$\tau_{\mathrm{BB}} = 2 e^2 /3 M c^3$ and $\Omega_{\mathrm{BB}}$ is a frequency cutoff.
This path-integral-based expression coincides completely with
the seminal results in Refs.~\citenum{FLO85,FLO88,FLO87} using the quantum
Langevin formalism. 
It also coincides with the result derived in Ref.~\citenum{BC91} using the standard path integral 
approach.
However, we need to note that in Ref.~\citenum{BC91}, the system, an electron, is interacting 
with its own radiation; here, by difference, we consider the system as being irradiated
by an external blackbody radiation such as sunlight or moonlight for $t>0$. 
This is precisely what allows us to separate the initial density operator of the system and the radiation. 
This natural emerging functional form of $J_{\mathrm{BB}}(\omega)$ reveals, from a
statistical viewpoint, the intrinsic non-Markovian character of the radiation \cite{FLO85,FLO88,FLO87,BC91}. 
This fact implies that
the two point correlation function, $\langle \hat{E}(t'') \hat{E}(t'') \rangle_{\mathrm{BB}}$,
of the electric field in Eq.~(\ref{equ:DefQuaElecField}) is not delta correlated.
From an optics point of view \cite{MW64a,*MW64b,*MW67} this means that
the blackbody radiation is coherent, although the coherence time is very short,
$\sim$1.3~fs at $T_{\mathrm{BB}} = 5900$~K (cf. Ref.~\citenum{PB13} and Chap.~13 in Ref.~\citenum{MW95}).

Once we have condensed the relevant information for the thermal bath and
the radiation field in the spectral densities Eqs.~(\ref{equ:JwTB}) and
(\ref{equ:JwBB}), we are in the position to define the various functions
entering into the functional actions $\Sigma[x,r,\bar{q}]$.

The kernels $K_{\mathrm{TB}}(s-\mathrm{i}\tau)$ and $k_{\mathrm{TB}}(\tau)$
are given by \cite{GSI88}
\begin{linenomath}
\begin{align}
\label{equ:KTB}
K_{\mathrm{TB}}(s-\mathrm{i}\tau) &= K^{\mathrm{re}}_{\mathrm{TB}}(s-\mathrm{i}\tau)
+ \mathrm{i} K^{\mathrm{im}}_{\mathrm{TB}}(s-\mathrm{i}\tau),
\\
\label{equ:kkTB}
k_{\mathrm{TB}}(\tau) &= \frac{m}{\hbar \beta_{\mathrm{TB}}}
\sum_{n=-\infty}^{\infty}\zeta_n(0) \exp(\mathrm{i} \nu_n \tau),
\end{align}
\end{linenomath}
with the Matsubara frequencies $\nu_n = 2\pi n/\hbar\beta_{\mathrm{TB}}$.
The kernel $K_{\mathrm{TB}}(s-\mathrm{i}\tau)$ contains the information of the 
thermal fluctuations due to TB and its influence in the lost of coherence as well 
as the in the decay of correlations between the system and the bath.
The kernel $k_{\mathrm{TB}}(\tau)$ contains the influence of the bath on the
thermal equilibrium state of the system S.
The real and imaginary parts of $K_{\mathrm{TB}}(s-\mathrm{i}\tau) $ are given
by
\begin{align}
 K^{\mathrm{re}}_{\mathrm{TB}}(s-\mathrm{i}\tau) &=
 \int\limits_{0}^{\infty} \frac{\mathrm{d} \omega}{\pi}
 J_{\mathrm{TB}}(\omega)\frac{\cosh[\omega(\frac{1}{2}\hbar\beta_{\mathrm{TB}})-\tau]}
                 {\sinh(\frac{1}{2}\omega\hbar\beta_{\mathrm{TB}})} \cos(\omega s)
\nonumber\\&
=\frac{m}{\hbar\beta_{\mathrm{TB}}}\sum_{n=-\infty}^{\infty}g_n(s)\exp(\mathrm{i}\nu_n \tau),
\\
 K^{\mathrm{im}}_{\mathrm{TB}}(s-\mathrm{i}\tau) &=
 -\int\limits_{0}^{\infty} \frac{\mathrm{d} \omega}{\pi}
 J_{\mathrm{TB}}(\omega)\frac{\sinh[\omega(\frac{1}{2}\hbar\beta_{\mathrm{TB}})-\tau]}
                 {\sinh(\frac{1}{2}\omega\hbar\beta_{\mathrm{TB}})} \sin(\omega s)
\nonumber \\&=\frac{m}{\hbar\beta_{\mathrm{TB}}}\sum_{n=-\infty}^{\infty}\mathrm{i}
f_n(s)\exp(\mathrm{i}\nu_n \tau).
\end{align}
while
$ \zeta_{n}(s) =\frac{1}{m}\int\limits_{0}^{\infty} \frac{\mathrm{d} \omega}{\pi}
               \frac{J_{\mathrm{TB}}(\omega)}{\omega}\
               \frac{2\nu_n^2}{\omega^2 + \nu_n^2}\cos(\omega s)$
or in terms of the damping kernel $\gamma_{\mathrm{TB}}(s)$ \cite{GSI88},
$ \zeta_{n}(s) = \frac{1}{2} |\nu_n|\int\limits_0^{\infty} \mathrm{d}u\,
\gamma_{\mathrm{TB}}(u) [\exp(-|\nu_s(s+u)|) + \exp(-|\nu_s(s-u)|) ]$.               
For the case when system-bath interactions are neglected, i.e., when there is no mixing 
between the imaginary time trajectory $\bar{q}(\tau)$ and the real time trajectory $x(s)$ in 
Eq.~(\ref{equ:GenSigma}), only the real part of $K_{\mathrm{TB}}(s)$ contributes to
the dynamics, which is consistent with calculations derived under the factorizing initial
condition assumption.

The functions $g_n(s)$ and $f_n(s)$ can be expressed in terms of the damping
kernels
\begin{equation}
\gamma_{\mathrm{TB},\mathrm{BB}}(s) 
=
\frac{2}{m} \int\limits_{0}^{\infty} \frac{\mathrm{d} \omega}{\pi}
               \frac{J_{\mathrm{TB},\mathrm{BB}}(\omega)}{\omega}\cos(\omega s),
\end{equation}
and $\zeta_n(s)$ as
$g_n(s) = \gamma_{\mathrm{TB}}(s)-\zeta_n(s)$ and
$f_n = -\frac{1}{\nu_n}\frac{\mathrm{d}}{\mathrm{d}s}\zeta_n(s)$.
The spectral density in Eq.~(\ref{equ:JwTB}) generates the damping
kernel
$\gamma(s) = \gamma_{\mathrm{TB}}
\Omega_{\mathrm{TB}} \exp(-\Omega_{\mathrm{TB}}|s|)$.
In the limit when the cutoff frequency $\Omega_{\mathrm{TB}}$ tends to infinity,
$\gamma(s) \rightarrow 2 \gamma_{\mathrm{TB}} \delta(s)$, which corresponds
to Markovian Ohmic dissipation. The spectral density in Eq.~(\ref{equ:JwBB})
generates
$
\gamma_{\mathrm{BB}}(s) =\tau_{\mathrm{BB}} \Omega_{\mathrm{BB}}^2
\left[2 \delta(s) - \Omega_{\mathrm{BB}} \exp(-\Omega_{\mathrm{BB}}|s|)\right].
$
Note that there is a fundamental limitation to the use of Eq. (\ref{equ:JwBB}).
That is, in the limit $\Omega_{\mathrm{BB}} \rightarrow\infty$, we get the
surprising result that
$\gamma_{\mathrm{BB}}(s) = 0$, i.e. no relaxation \cite{FO91,FO98}.
This corresponds to  the point-electron limit
[$f_k = \Omega_{\mathrm{BB}}^2/(\Omega_{\mathrm{BB}}^2 + \omega_k^2) = 1$
in Eq.~(\ref{equ:AVPBBR})] and is unphysical  because even for the electron,
$\Omega_{\mathrm{BB}}$ remains finite, although large.
According to Refs.~\citenum{FO91,FO98}, there is a natural upper value given by
$\Omega_{\mathrm{BB}} = {\tau_{\mathrm{BB}}}^{-1}$, which corresponds to
two-thirds of the time for a photon to traverse the classical electron radius
($r_\mathrm{cl}^{\mathrm{e}} = 2.818\times10^{-15}$m).
Beyond this natural limit, causality is violated \cite{FO91} and the bare mass $m$
takes negative values \cite{FO91}.

Finally, the kernel $K_{\mathrm{BB}}(s)$ is given by
\begin{equation}
\label{equ:KBB}
 K_{\mathrm{BB}}(s) = \int\limits_{0}^{\infty} \frac{\mathrm{d} \omega}{\pi}
               J_{\mathrm{BB}}(\omega)
               \coth\left(\frac{\omega \hbar\beta_{\mathrm{BB}}}{2}\right)\cos(\omega s).
\end{equation}
This kernel is responsible for the decoherence due to thermal fluctuations induced 
by the blackbody radiation, while the kernels 
$\eta_{\mathrm{TB}}(s)=m \gamma_{\mathrm{TB}}(s)$ and
$\eta_{\mathrm{BB}}(s)=  m \gamma_{\mathrm{BB}}(s)$ in Eq.~(\ref{equ:GenSigma})
induce the relaxation process.

\subsection{Explicit calculation of the propagating function}
The explicit calculation of the propagating function demands  evaluating  the path integral  in 
Eq.~(\ref{equ:GenPropFunct}).
Since, the system is linear, the path integral can be performed by evaluating
the action in Eq.~(\ref{equ:GenSigma}) along its stationary trajectories and
condensing the effect of the fluctuations in a global time dependent factor
\cite{GSI88,Wei08}.
The extremum of the action for imaginary time is given by  
\begin{equation}
m \ddot{\bar{q}} - m \omega_0^2 \bar{q} -
\int\limits_0^{\hbar \beta_{\mathrm{TB}}}
\mathrm{d} \sigma k_{\mathrm{TB}}(\tau -\sigma) \bar{q}(\sigma)
= -\mathrm{i}\int\limits_0^t \mathrm{d} s K_{\mathrm{TB}}^*(s-\mathrm{i}\tau)x(s),
\end{equation}
where we can see how the dynamics in real time, represented by $x(s)$, drives
the system-bath correlations, by driving the imaginary time path $q(\tau)$ in a 
non-local way.
For real time, the  action is stationary along
\begin{align} 
\begin{split}
&\hspace{-0.5cm}m \ddot{r} + m \omega(t)^2 r -  E_{\mathrm{L}}(t)
+\frac{\mathrm{d}}{\mathrm{d} s}\int\limits_0^s \mathrm{d} u \,\eta(s-u)r(u)
\\
&\hspace{-0.5cm}= r'\eta_{\mathrm{BB}}(s) + \mathrm{i}\int\limits_0^t \mathrm{d} u K(s-u)x(u)
+ \hspace{-0.25cm}\int\limits_0^{\hbar \beta_{\mathrm{TB}}}  \mathrm{d} \hspace{-0.0125cm} \tau
K_{\mathrm{TB}}^*(s - \mathrm{i} \tau)\bar{q}(\tau),
\end{split}
\\
\begin{split}
m \ddot{x} &+ m \omega(t)^2 x 
- \frac{\mathrm{d}}{\mathrm{d} s}\int\limits_s^t \mathrm{d} u \,\eta(s-u)x(u)=0,
\end{split}
\end{align}
where we have defined 
\begin{equation*}
\eta(s) = \eta_{\mathrm{TB}}(s) + \eta_{\mathrm{BB}}(s), \quad 
K(s) = K^{\mathrm{re}}_{\mathrm{TB}}(s) + K_{\mathrm{BB}}(s).
\end{equation*}
The term $ r'\eta_{\mathrm{BB}}(s)$ appears here as a consequence of the sudden
turn-on of the blackbody radiation. Since, we assume that the parametric driving,
as well as the laser field and the blackbody radiation act after $t=0$, the equilibrium
state of our system coincides with the one derived in Ref.~\citenum{GSI88}, so we
 need to focus only on the evaluation of the real part of the action.

The real part of the action is stationary along the solution to the
equation of motion
\begin{linenomath}
\begin{align}
\begin{split}
\label{equ:FinalEOMr}
m \ddot{r} &+ m \frac{\mathrm{d}}{\mathrm{d} s}\int\limits_0^s \mathrm{d} u \,\gamma(s-u)r(u)
+ m \omega(t)^2 r
\\
&=\bar{E}_{\mathrm{L}}(s) + \mathrm{i} m \int\limits_0^t \mathrm{d} u R(s-u)x(u),
\end{split}
\\
\label{equ:FinalEOMx}
m \ddot{x}  &- m \frac{\mathrm{d}}{\mathrm{d} s}\int\limits_s^t \mathrm{d} u \,\gamma(s-u)x(u)
+ m \omega(t)^2 x = 0,
\end{align}
\end{linenomath}
where we have defined
$\bar{E}_{\mathrm{L}}(s) =
E_{\mathrm{L}}(s) + r'\eta_{\mathrm{BB}}(s) + m [\bar{x} C_1(s) -
\mathrm{i} \bar{x} C_2(s)]$ with $\bar{r} = (\bar{q}+\bar{q}')/2$ and
$\bar{x} = \bar{q}-\bar{q}'$. Additionally, we have defined
$\eta(s) = m \gamma(s)$ and 
\begin{equation*}
R(s,u) = R_{\mathrm{TB}}(s,u) +  K(s-u)/m
\end{equation*}
 with
\begin{align}
\begin{split}
 R_{\mathrm{TB}}&(s,u) = -\Lambda_{\mathrm{TB}} C_{1}(s)C_{1}(u)
\\&
+ \frac{1}{\hbar\beta_{\mathrm{TB}}}
\sum_{n=-\infty}^{\infty} u_{n}[g_{n}(s) g_{n}(u) - f_{n}(s) f_{n}(u)],
\end{split}
\\
 C_{1}(s) &= \frac{1}{\hbar\beta_{\mathrm{TB}}\Lambda_{\mathrm{TB}}}
 \sum_{n=-\infty}^{\infty} u_{n} g_{n}(s),
\\
 C_{2}(s) &= \frac{1}{\hbar\beta_{\mathrm{TB}}}\sum_{n=-\infty}^{\infty} u_{n} \nu_{n} f_{n}(s),
\end{align}
where $ \Lambda_{\mathrm{TB}} =\frac{1}{\hbar\beta_{\mathrm{TB}}}
\sum_{n=-\infty}^{\infty} u_{n}$ can be related to the second moment of the position
of the system at equilibrium with TB, 
$\langle q^2 \rangle_{\mathrm{equil.}} = (\hbar/m)\Lambda_{\mathrm{TB}}$, and 
$u_{n} = (\omega_0^2 + \nu_n^2 + \zeta_n)^{-1}$.

Since, for a harmonic potential, the functional action $\Sigma [x,r,\bar{q}_x]$
can be evaluated using only the real part of the trajectories $r(s)$ and $x(s)$
\cite{HA85,SG86,GSI88,PID10}, we  need to solve only for the real part of Eq.
(\ref{equ:FinalEOMr}) and (\ref{equ:FinalEOMx}).
Due to the linear character of (\ref{equ:FinalEOMr}), the solution to the
homogeneous part can be written as
\begin{equation}
\begin{split}
r(s) &= r'' \frac{\phi_1(s)}{\phi_1(t)} + r' \left(\phi_2(s) - 
\frac{\phi_2(t)}{\phi_1(t)} \phi_1(s) \right),
\end{split}
\end{equation}
where $\phi_1(s)$ is the fundamental solution for $r(0)=0$ and $\dot{r}(0)=1$,
while $\phi_2(s)$ is the fundamental solution for $r(0)=1$ and $\dot{r}(0)=0$.
Thus, for Eqs. (\ref{equ:FinalEOMr}) and (\ref{equ:FinalEOMx}) we have
\begin{align}
\label{equ:Finar}
\begin{split}
 r^{\mathrm{re}}(s) &= r'' \frac{\phi_1(s)}{\phi_1(t)} + r'\left(\phi_2(s) - \frac{\phi_2(t)}{\phi_1(t)} \phi_1(s) \right)
\\       &+ \frac{1}{m}\int_0^s \mathrm{d}u \phi_1(s-u) \bar{F}'(u)
           - \frac{1}{m}\frac{\phi_1(s)}{\phi_1(t)}\int_0^t \mathrm{d}u \phi_1(t-u) \bar{F}'(u).
\end{split}
\end{align}
Since $\bar{F}'(s)$ contains the term induced by the sudden coupling to the radiation, 
$r'\eta_{\mathrm{BB}}(s)$, we can see that $r(s)$ is driven by this sudden turn on.

Following a similar procedure for $x(s)$, we get
\begin{align}
\label{equ:Finax}
\begin{split}
 x(s) &= x'' \frac{\varphi_1(s)}{\varphi_1(t)} + x' \left(\varphi_2(s) -
 \frac{\varphi_2(t)}{\varphi_1(t)} \varphi_1(s) \right),
\end{split}
\end{align}
where $\varphi_1(s)$ is the fundamental solution for $x(0)=0$ and $\dot{x}(0)=1$,
while $\varphi_2(s)$ is the fundamental solution for $x(0)=1$ and $\dot{x}(0)=0$.

For the particular case of no frequency modulation, $\phi_1(s)$, $\phi_2(s)$,
$\varphi_1(s)$ and $\varphi_2(s)$ can be derived from standard Laplace techniques
\cite{GSI88}.
For Markovian dissipation, $\Omega_{\mathrm{TB}}\rightarrow \infty$ in Eq.~(\ref{equ:JwTB}),
and harmonic modulation of the frequency, $\phi_1(s)$, $\phi_2(s)$, $\varphi_1(s)$ and $\varphi_2(s)$
are related to the Mathieu functions \cite{ZH95,GPZ10}.
For more general cases, these functions must be calculated numerically.
However, the functional form of Eqs. (\ref{equ:Finar}) and (\ref{equ:Finax})
is very convenient of the subsequent analytical calculations.
For further convenience we define
\begin{align}
\label{equ:v1v2}
v_1(t,s) &= \varphi_2(s) -  \frac{\varphi_2(t)}{\varphi_1(t)} \varphi_1(s),
\quad
v_2(t,s) =\frac{\varphi_1(s)}{\varphi_1(t)} ,
\\
\label{equ:u1u2}
u_1(t,s) &= \phi_2(s) -  \frac{\phi_2(t)}{\phi_1(t)} \phi_1(s),
\quad
u_2(t,s) =\frac{\phi_1(s)}{\phi_1(t)}.
\end{align}

The influence functional in Eq.~(\ref{equ:GenPropFunct}) can now be
rewritten as
\begin{equation}
\label{equ:GenPropFunctFinal}
\begin{split}
J(r'',x'',t;r',x',0;\bar{r},\bar{x})  =
\frac{1}{N(t)}
\exp\left(\frac{\mathrm{i}}{\hbar} \Sigma[r,x,\bar{r},\bar{x}]\right),
\end{split}
\end{equation}
where $N(t)$ is a normalization factor given by
$N(t) = 2\pi \hbar \frac{1}{m} |\dot{u}_2(t,0)
|\left(2\pi \hbar \frac{1}{m}\Lambda_{\mathrm{TB}}\right)^{1/2}$.
After evaluating Eq. (\ref{equ:GenSigma}) along $r^{\mathrm{re}}(s)$ and $x(s)$, we get
\begin{widetext}
\begin{equation}
\begin{split}
\label{equ:FinalSigma}
\Sigma& [r'',x'',r',x',\bar{r},\bar{x}]=
\\
& \mathrm{i}m \left( \frac{1}{2\Lambda_{\mathrm{TB}}}\bar{r}^2
+ \frac{\Omega_{\mathrm{TB}}}{2}\bar{x}^2\right)
+ m \left[x'' r'' \dot{u}_2(t,t) + x' r' \dot{u}_1(t,0) \right]
-m\left[x' r'' \dot{u}_2(t,0) - x'' r' \dot{u}_1(t,t) \right]
\\
& + m\int\limits_0^{t} \mathrm{d}s
\left[x' v_1(t,s)+ x'' v_2(t,s) \right]
\left[\frac{1}{m}E_{\mathrm{L}}(s) + r' \gamma_{\mathrm{BB}}(s) +
\bar{r} C_1(s) - \mathrm{i}\bar{x}C_2(s)\right]
\\
\phantom{\Sigma [r'',x'',r',x',\bar{r},\bar{x}]}\hspace{-0.75cm}
&+\frac{\mathrm{i}}{2}m {x'}^2 \int_0^t \mathrm{d}s  \int_0^t \mathrm{d}u
R(s,u)  v_1(t,s) v_1(t,u)
+\frac{\mathrm{i}}{2}m x' x'' \int_0^t \mathrm{d}s  \int_0^t \mathrm{d}u
R(s,u) v_1(t,s) v_2(t,u)
\\
&+\frac{\mathrm{i}}{2}m x'' x'\int_0^t \mathrm{d}s  \int_0^t \mathrm{d}u
R(s,u) v_2(t,s) v_1(t,u)
+\frac{\mathrm{i}}{2}m {x''}^2 \int_0^t \mathrm{d}s  \int_0^t \mathrm{d}u
R(s,u)  v_2(t,s) v_2(t,u),
\end{split}
\end{equation}
\end{widetext}
where $\Omega_{\mathrm{TB}} =\frac{1}{\hbar\beta_{\mathrm{TB}}}
\sum_{n=-\infty}^{\infty} u_{n}(\omega_0^2 + \zeta_n)$ can be related
to the second moment of the momentum, 
$\langle p^2 \rangle_{\mathrm{equil.}} = \hbar m\Omega_{\mathrm{TB}}$,
at equilibrium with TB.
The first term in Eq.~(\ref{equ:FinalSigma}) containing $\Lambda_{\mathrm{TB}}$
and $\Omega_{\mathrm{TB}}$ can be associated to the thermal equilibrium state
influenced by the presence of the thermal bath TB.
These results provide the general expression for the influence functional.
\subsection{Limiting cases}
The general result in Eq.  (\ref{equ:GenPropFunctFinal}) 
includes, and agrees with, several limiting cases. These include:
\begin{enumerate}[i]
\item \emph{In absence of blackbody radiation and with no parametric
modulation of the frequency}, Eq.~(\ref{equ:GenPropFunctFinal}) reduces to the result in
Ref.~\citenum{GSI88}.

\item \emph{In absence of the thermal bath and for no laser field and
no parametric modulation of the frequency}, Eq.~(\ref{equ:GenPropFunctFinal}) is the
formal path integral equivalent of Refs.~\citenum{FLO85,FLO88} (based on the quantum
Langevin equation formalism).

\item \emph{In absence of the thermal bath and with no parametric
modulation of the frequency}, Eq.~(\ref{equ:GenPropFunctFinal}) is the formal path integral
equivalent of Refs.~\citenum{FLO85,FLO88} to the result in Refs.~\citenum{LFO90,LFO90b}.

\item \emph{In absence of blackbody radiation, for no laser field,
harmonic modulation of the frequency and factorized initial conditions},
Eq.~(\ref{equ:GenPropFunctFinal}) reduces to Ref.~\citenum{ZH95} (see also Ref.~\citenum{GPZ10}).

\item \emph{In absence of  blackbody radiation, for no laser field,
no modulation of the frequency and for factorized initial conditions},
Eq.~(\ref{equ:GenPropFunctFinal}) reduces to Refs.~\citenum{CL83,GWT84} (see also Ref.~\citenum{PID10}
for a description in terms of the Wigner function and Ref.~\citenum{HPZ92} for a master equation 
approach).
\end{enumerate}

\subsection{Explicit form of the propagating function}
For an initial thermal state, i.e.  $\lambda(q_{+}',{\bar q}'',q_{-}',{\bar q}') = \delta(q_{+}'-{\bar q}'')
\delta(q_{-}'-{\bar q}')$ in Eq.~(\ref{equ:initialStateGen}) and correspondingly $\bar{r}=r'$ and
$\bar{x} = x'$, the influence functional in Eq.~(\ref{equ:GenPropFunctFinal}) with the function
phase given in Eq.~(\ref{equ:FinalSigma}) can be written in the very compact form
\begin{align}
\label{equ:InfFuncMatForm}
\begin{split}
J&(r'',x'',t;,r',x',0;r',x') = \frac{1}{N(t)}
\\&\times
\exp\left\{-\frac{1}{2} \mathbf{x}^{\mathrm{T}} \mathsf{A}(t) \mathbf{x}
+ E_{\mathrm{L}}^+(s) x' +  E_{\mathrm{L}}^-(s) x'' \right\}
\end{split}
\end{align}
where
$ \mathbf{x}^{\mathrm{T}} \mathsf{A}(t) \mathbf{x}=
 \mathbf{x}_f^{\mathrm{T}} \mathsf{A}_f(t) \mathbf{x}_f+
\mathbf{x}_i^{\mathrm{T}} \mathsf{A}_i(t) \mathbf{x}_i
+2\mathbf{B} \mathbf{x}_i$
where $\mathbf{x} = (x'', r'',x',r')$, $\mathbf{x}_{i} = (x', r')$,
 $\mathbf{x}_f = (x'', r'')$, 
 $\mathbf{B} = (\mathsf{A}_{13} x'' + \mathsf{A}_{23} r'', \mathsf{A}_{14} x'')$
with the time dependent matrix $\mathsf{A}(t)$ given by
\begin{widetext}
\begin{equation}
\mathsf{A}(t) = \frac{m}{\hbar}\left(
\begin{array}{cccc}
R_{22}(t) & -\mathrm{i} \dot{u}_2(t,t)& -C_2^-(t) + R_{12}(t)&
-\mathrm{i} \dot{u}_1(t,t) - \mathrm{i} \tilde{C}_1^-(t)
\\
-\mathrm{i}\dot{u}_2(t,t)& 0 & \mathrm{i} \dot{u}_2(t,0) & 0
\\
-C_2^-(t) + R_{12}(t) & \mathrm{i} \dot{u}_2(t,0)  &
\Omega_{\mathrm{TB}} - 2 C_2^+(t) + R_{11}(t)&
-\mathrm{i} \dot{u}_1(t,0) - \mathrm{i} \tilde{C}_1^+(t)
\\
-\mathrm{i} \dot{u}_1(t,t) - \mathrm{i} \tilde{C}_1^-(t)& 0 &
-\mathrm{i} \dot{u}_1(t,0) - \mathrm{i} \tilde{C}_1^+(t)&
1/\Lambda_{\mathrm{TB}}
\end{array}
\right),
\end{equation}
\end{widetext}
being
\begin{linenomath}
\begin{equation}
\label{equ:Cjpm}
C_j^+(t) =
\int\limits_0^t  \mathrm{d}s  C_j(s) v_1(t,s),
\quad
C_j^-(t) =
\int\limits_0^t \mathrm{d}s C_j(s) v_2(t,s).
\end{equation}
The $C_j^{\pm}$s functions account for the influence of the initial correlations between 
the system and the bath on the system dynamics. $\tilde{C}_1^{\pm}(t)$ is obtained by replacing 
$C_1(t) \rightarrow C_1(t) +  \gamma_{\mathrm{BB}}(t) $ in the definition of 
$C_1^{\pm}(t)$ in Eq.~(\ref{equ:Cjpm}) and contains the effect of the turn on of the
interaction with the blackbody radiation.
The effects of the laser field $E_{\mathrm{L}}$ on the dynamics are encoded in 
\begin{equation}
\label{equ:Epm}
E_{\mathrm{L}}^+(t) =
\int\limits_0^t  \mathrm{d}s  E_{\mathrm{L}}(s) v_1(t,s),
\quad
E_{\mathrm{L}}^-(t) =
\int\limits_0^t \mathrm{d}s E_{\mathrm{L}}(s)v_2(t,s).
\end{equation}
The decoherence dynamics is governed by the $R_{ij}(t)$ functions given by
\begin{equation}
\label{equ:Rij}
R_{ij}(t) =
\int_0^t \mathrm{d}s  \int_0^t \mathrm{d}u R(s,u)
v_i(t,s) v_j(t,s).
\end{equation}
\end{linenomath}

Once we have completely characterized the functional form of the propagating
function, we proceed in the next sections to derive explicit expressions for the
covariance matrix elements of the quantum fluctuations and the time evolution
of the reduced density-operator.

\section{Time evolution of the covariance matrix elements}
Due to the linear nature of the system under consideration, the dynamics as well as the statistical
properties can be characterized in terms of the first and second moments \cite{GWT84},
defined as
$\langle f(q(t)) \rangle = \int \mathrm{d} r''
f(r'') \rho_{\mathrm{S}}(r'',x''=0,t)$, or in terms of the propagating function in
Eq.~(\ref{equ:InfFuncMatForm}) by,
\begin{equation}
\langle f(q(t)) \rangle = \int \mathrm{d} r'' \mathrm{d} r' \mathrm{d} x'
f(r'') J(r'',x''=0,r',x',r',x').
\end{equation}
So, we can now make use of the explicit form of the propagating 
function in Eq.~(\ref{equ:InfFuncMatForm}).
Note that the propagating function in Eq.~(\ref{equ:InfFuncMatForm})
also contains  the initial thermal state $\hat{\rho}_{\beta,\mathrm{S}}$.
These moments define the variances or dispersion relations
\begin{align}
\label{equ:sigmaqq}
\sigma_{qq}(t) &= \langle q^2(t) \rangle - \langle q(t) \rangle^2,
\\
\label{equ:sigmaqp}
\sigma_{qp}(t) &= \frac{1}{2}\langle q(t)p(t) + p(t)q(t)\rangle
- \langle q(t) \rangle\langle p(t) \rangle,
\\
\label{equ:sigmapp}
\sigma_{pp}(t) &= \langle p^2(t) \rangle - \langle p(t) \rangle^2,
\end{align}
which will be used in order to express explicitly  the time evolution of the
density matrix in Eq.~(\ref{equ:DenMatSFinal}).

The first moments are determined by
\begin{align}
\label{equ:q}
\langle q(t) \rangle &= \frac{1}{m \dot{u}_2(t,0)}\int_0^t \textrm{d}s\, v_1(t,s)E_{\mathrm{L}}(s)
\\
\label{equ:p}
\langle p(t) \rangle &=
\int_0^t \textrm{d}s \left[v_1(t,s) \frac{\dot{u}_2(t,t)}{\dot{u}_2(t,0)}
+ v_2(t,s)\right] E_{\mathrm{L}}(s)
\end{align}
where the overdot denotes the derivative with respect to $s$, i.e., 
$\dot{u}_2(t,0) = \partial u_2(t,s)/\partial s|_{s=0}$.

The second moments are given by
\begin{align}
\begin{split}
\label{equ:qq}
\langle q^2(t) \rangle &= \langle q(t) \rangle^2
+
\frac{\hbar^2}{m^2 \dot{u}^2_2(t,0)}
\left[M_{11}(t) -\frac{\hbar}{m}{M_{12}(t)}^2\right] ,
\end{split}
\\
\begin{split}
\label{equ:pq}
\frac{1}{2}\langle pq + qp \rangle &= m \langle q^2(t) \rangle \dot{u}_2(t,t)
+ \frac{\hbar}{\dot{u}_2(t,0)} \left[R_{12}(t) - C_2^-(t) \right]
\\ &+ \mathrm{i} \frac{\hbar^2\Lambda_{\mathrm{TB}}}{m \dot{u}_2(t,0)}
\left[\tilde{C}_1^-(t) + \dot{u}_1(t,t)\right]M_{12}(t)
\\
&+  \langle q(t) \rangle \int_0^t \mathrm{d}s v_2(t,s)E_{\mathrm{L}}(s),
\end{split}
\\
\begin{split}
\label{equ:pp}
\langle p^2(t) \rangle &= \hbar m R_{22}(t)
+ \hbar m \Lambda_{\mathrm{TB}} \left[\tilde{C}_1^-(t) +\dot{u}_1(t,t)\right]^2
\\ &
- m^2 \dot{u}^2_2(t,t)\langle q^2(t) \rangle
+m\dot{u}_2(t,0)\langle pq + qp \rangle
\\
&+ \left[\int_0^t \hspace{-0.1cm}\mathrm{d}s v_2(t,s)E_{\mathrm{L}}(s)\right]^2,
\end{split}
\end{align}
where
\begin{equation}
\mathsf{M} = \frac{m}{\hbar}
\left(
\begin{tabular}{cc}
$\Omega_{\mathrm{TB}} - 2C_2^+(t) + R_{11}$ &
$-\mathrm{i}\left[\dot{u}_1(t,0) + \tilde{C}_1^+(t)\right]$
\\
$-\mathrm{i}\left[\dot{u}_1(t,0) + \tilde{C}_1^+(t)\right]$ &
$1/\Lambda_{\mathrm{TB}}$
\end{tabular}
\right).
\end{equation}

In the absence of the parametric driving, the blackbody radiation and the laser field,
Eqs.~(\ref{equ:qq})-(\ref{equ:pp}) are time-independent and coincide with the
expressions (6.62)-(6.64) in Ref.~\citenum{GSI88}, i.e.,
$\langle q^2(t) \rangle = (\hbar/m)\Lambda_{\mathrm{TB}}$,
$\langle pq + qp \rangle = 0$ and
$\langle p^2(t) \rangle = \hbar m \Omega_{\mathrm{TB}} $.

\section{Time evolution of the reduced density-operator}
If one is interested in the reduced-density operator itself, it can be written in terms of the second moments
as
\begin{equation}
\label{equ:DenMatSFinal}
\begin{split}
\bok{r''}{\hat{\rho}_{\mathrm{S}}(t)}{x''} &= \left(2\pi\sigma_{qq}(t)\right)^{-1/2}
\exp\left[-\frac{1}{2\sigma_{qq}(t)}\left[r'' - \langle q(t) \rangle\right]^2
\right. \\
&- \left. \frac{1}{2\hbar^2} \left(\sigma_{pp}(t)
- \frac{\sigma_{pq}(t)^2}{\sigma_{qq}(t)} \right) {x}''^2
\right.
\\
&+ \left.\frac{\iu}{\hbar}\left\{ \langle p(t) \rangle +
 \frac{\sigma_{pq}(t)^2}{\sigma_{qq}(t)}
 \left(r'' - \langle q(t) \rangle \right)\right\}x''\right].
\end{split}
\end{equation}
In summary, to obtain the time evolution of the reduced density operator, we proceed
as follows:
\begin{enumerate}[i]
\item One first specifies the spectral density $J_{\mathrm{TB}}(\omega)$ 
to obtain the function 
describing the modulating force in Eq.~(\ref{equ:HamSysTBBBb}).
This permits us to obtain 
the fundamental solutions $\phi_{1,2}(s)$ and $\varphi_{1,2}(s)$ 
in Eq.~(\ref{equ:Finar}) 
and  Eq.~(\ref{equ:Finax}), respectively.
\item One then calculates  
the kernels $K_{\mathrm{TB}}(s)$, $k(\tau)$ and $K_{\mathrm{BB}}(s)$
defined in Eqs.~(\ref{equ:KTB}), (\ref{equ:kkTB}) and (\ref{equ:KBB}).
\item With the fundamental solutions  obtained, and all the kernels calculated, 
we calculate the auxiliary functions $v_{1,2}(t,s)$ and $u_{1,2}(t,s)$ given in Eqs.~(\ref{equ:v1v2}) 
and (\ref{equ:v1v2}), and subsequently the functions $C_j^{\pm}(t)$, $E^{\pm}(t)$ and $R_{ij}(t)$ 
defined by Eqs.~(\ref{equ:Cjpm}), (\ref{equ:Epm}) and (\ref{equ:Rij}), respectively.
\item One then calculates the first and second moments given in
Eqs.~(\ref{equ:q}-\ref{equ:pp}), and subsequently the dispersion relations in 
Eqs.~(\ref{equ:sigmaqq}-\ref{equ:sigmapp}) and system dynamics via Eq. (\ref{equ:DenMatSFinal}). 
\end{enumerate}

This brief prescription concludes our completely formal and approximation-free
treatment.
Note that we have successfully applied the method to a number of cases,
some of which are reported elsewhere \cite{PB13,PB13c}.
\section{Concluding Remarks}
As discussed in the introduction, the results derived here can be used to study a wide variety of 
problems, e.g., the incoherent \cite{PB13} or coherent \cite{PB13c} excitation of open quantum
systems in order to provide physical insight into the role of coherences detected
in photosynthetic light-harvesting complexes (for a review in the subject see Ref.~\citenum{PB12c}).
In doing so, we need to translate the propagating function in Eq.~(\ref{equ:InfFuncMatForm})
into the energy basis in order to identify the incoherent/coherent nature of the
excitation. 

Our results allow us to directly study the possibility of environmentally assisted one-photon phase control 
\cite{SAB10,PYB13} provided by the fact that the initial equilibrium density matrix deviates from the canonical
distribution. In this respect, we 
eliminate the incoherent radiation and the frequency modulation 
contributions and 
focus on how the phase information encoded in $E_{\mathrm{L}}(t)$ can be used to manipulate the populations 
of the oscillator \cite{PB13c}.

Additionally, the analytic closed expression could be useful in understanding the
delicate balance between dissipation and driving under non-Markovian evolution
that has been pointed out in Ref.~\citenum{SN&11} in the context of optimal control theory
and cooling of nano-mechanical resonators.
In particular, the optimal cooling protocol addressed in Ref.~\citenum{SN&11} by means of numerical
techniques, can be analyzed in great detail from the second moments derived in 
Eqs.~(\ref{equ:qq})--(\ref{equ:pp}) and the theory of variational calculus. 

In Ref. ~\citenum{GPZ10}, it was established that the usual quantum limit, $\hbar \omega / k_{\mathrm{B}}T >1$, 
needs to be reformulated for out-of-equilibrium systems. However, in that work a Markovian Ohmic
spectral density, $\Omega_{\mathrm{TB}}\rightarrow \infty$ in Eq.~(\ref{equ:JwTB}), was used, disregarding 
in this way the dynamics during non-Markovian time scales. 
The physical system considered in Ref.~\citenum{GPZ10} consisted of two identical harmonic oscillators
with time-depend coupling, in the normal mode description we get two independent parametric 
oscillators. So, under the same circumstances considered in Ref.~\citenum{GPZ10}, the results derived here
allow us to explore the limit for the presence of quantum features in 
non-Markovian-driven-open-quantum systems, which is of great importance in, e.g., quantum 
statistical mechanics or control theory.

\begin{acknowledgements}
LAP acknowledges discussions with Gert-Ludwig Ingold with pleasure.
This work was supported by the US Air Force Office of Scientific Research under contract
number FA9550-10-1-0260, by \emph{Comit\'e para el Desarrollo de la Investigaci\'on} 
--CODI-- of Universidad de Antioquia, Colombia, under contract number E01651 and by 
the \textit{Colombian Institute for the Science and Technology Development} --COLCIENCIAS--
under the contract number 111556934912 
\end{acknowledgements}

\bibliography{jmpqho}

\begin{thebibliography}{58}%
\makeatletter
\providecommand \@ifxundefined [1]{%
 \@ifx{#1\undefined}
}%
\providecommand \@ifnum [1]{%
 \ifnum #1\expandafter \@firstoftwo
 \else \expandafter \@secondoftwo
 \fi
}%
\providecommand \@ifx [1]{%
 \ifx #1\expandafter \@firstoftwo
 \else \expandafter \@secondoftwo
 \fi
}%
\providecommand \natexlab [1]{#1}%
\providecommand \enquote  [1]{``#1''}%
\providecommand \bibnamefont  [1]{#1}%
\providecommand \bibfnamefont [1]{#1}%
\providecommand \citenamefont [1]{#1}%
\providecommand \href@noop [0]{\@secondoftwo}%
\providecommand \href [0]{\begingroup \@sanitize@url \@href}%
\providecommand \@href[1]{\@@startlink{#1}\@@href}%
\providecommand \@@href[1]{\endgroup#1\@@endlink}%
\providecommand \@sanitize@url [0]{\catcode `\\12\catcode `\$12\catcode
  `\&12\catcode `\#12\catcode `\^12\catcode `\_12\catcode `\%12\relax}%
\providecommand \@@startlink[1]{}%
\providecommand \@@endlink[0]{}%
\providecommand \url  [0]{\begingroup\@sanitize@url \@url }%
\providecommand \@url [1]{\endgroup\@href {#1}{\urlprefix }}%
\providecommand \urlprefix  [0]{URL }%
\providecommand \Eprint [0]{\href }%
\providecommand \doibase [0]{http://dx.doi.org/}%
\providecommand \selectlanguage [0]{\@gobble}%
\providecommand \bibinfo  [0]{\@secondoftwo}%
\providecommand \bibfield  [0]{\@secondoftwo}%
\providecommand \translation [1]{[#1]}%
\providecommand \BibitemOpen [0]{}%
\providecommand \bibitemStop [0]{}%
\providecommand \bibitemNoStop [0]{.\EOS\space}%
\providecommand \EOS [0]{\spacefactor3000\relax}%
\providecommand \BibitemShut  [1]{\csname bibitem#1\endcsname}%
\let\auto@bib@innerbib\@empty
\bibitem [{\citenamefont {Magalinski{\u\i}}(1959)}]{Mag59}%
  \BibitemOpen
  \bibfield  {author} {\bibinfo {author} {\bibfnamefont {V.~B.}\ \bibnamefont
  {Magalinski{\u\i}}},\ }\href@noop {} {\bibfield  {journal} {\bibinfo
  {journal} {Zh. Eksp. Teor. Fiz.}\ }\textbf {\bibinfo {volume} {36}},\
  \bibinfo {pages} {1942} (\bibinfo {year} {1959})}\BibitemShut {NoStop}%
\bibitem [{\citenamefont {Feynman}\ \emph {et~al.}(1963)\citenamefont {Feynman}
  \emph {et~al.}}]{FV63}%
  \BibitemOpen
  \bibfield  {author} {\bibinfo {author} {\bibfnamefont {R.~P.}\ \bibnamefont
  {Feynman}} \emph {et~al.},\ }\href {\doibase 10.1016/0003-4916(63)90068-X}
  {\bibfield  {journal} {\bibinfo  {journal} {Annals of Physics}\ }\textbf
  {\bibinfo {volume} {24}},\ \bibinfo {pages} {118} (\bibinfo {year}
  {1963})}\BibitemShut {NoStop}%
\bibitem [{\citenamefont {Ullersma}(1966{\natexlab{a}})}]{Ull66}%
  \BibitemOpen
  \bibfield  {author} {\bibinfo {author} {\bibfnamefont {P.}~\bibnamefont
  {Ullersma}},\ }\href {\doibase 10.1016/0031-8914(66)90102-9} {\bibfield
  {journal} {\bibinfo  {journal} {Physica}\ }\textbf {\bibinfo {volume} {32}},\
  \bibinfo {pages} {27} (\bibinfo {year} {1966}{\natexlab{a}})}\BibitemShut
  {NoStop}%
\bibitem [{\citenamefont {Ullersma}(1966{\natexlab{b}})}]{Ull66b}%
  \BibitemOpen
  \bibfield  {author} {\bibinfo {author} {\bibfnamefont {P.}~\bibnamefont
  {Ullersma}},\ }\href {\doibase 10.1016/0031-8914(66)90103-0} {\bibfield
  {journal} {\bibinfo  {journal} {Physica}\ }\textbf {\bibinfo {volume} {32}},\
  \bibinfo {pages} {56} (\bibinfo {year} {1966}{\natexlab{b}})}\BibitemShut
  {NoStop}%
\bibitem [{\citenamefont {Ullersma}(1966{\natexlab{c}})}]{Ull66c}%
  \BibitemOpen
  \bibfield  {author} {\bibinfo {author} {\bibfnamefont {P.}~\bibnamefont
  {Ullersma}},\ }\href {\doibase 10.1016/0031-8914(66)90104-2} {\bibfield
  {journal} {\bibinfo  {journal} {Physica}\ }\textbf {\bibinfo {volume} {32}},\
  \bibinfo {pages} {74} (\bibinfo {year} {1966}{\natexlab{c}})}\BibitemShut
  {NoStop}%
\bibitem [{\citenamefont {Ullersma}(1966{\natexlab{d}})}]{Ull66d}%
  \BibitemOpen
  \bibfield  {author} {\bibinfo {author} {\bibfnamefont {P.}~\bibnamefont
  {Ullersma}},\ }\href {\doibase 10.1016/0031-8914(66)90105-4} {\bibfield
  {journal} {\bibinfo  {journal} {Physica}\ }\textbf {\bibinfo {volume} {32}},\
  \bibinfo {pages} {90} (\bibinfo {year} {1966}{\natexlab{d}})}\BibitemShut
  {NoStop}%
\bibitem [{\citenamefont {Caldeira}\ and\ \citenamefont
  {Leggett}(1983)}]{CL83}%
  \BibitemOpen
  \bibfield  {author} {\bibinfo {author} {\bibfnamefont {A.~O.}\ \bibnamefont
  {Caldeira}}\ and\ \bibinfo {author} {\bibfnamefont {A.~L.}\ \bibnamefont
  {Leggett}},\ }\href {\doibase 10.1016/0378-4371(83)90013-4} {\bibfield
  {journal} {\bibinfo  {journal} {Physica A}\ }\textbf {\bibinfo {volume}
  {121}},\ \bibinfo {pages} {587} (\bibinfo {year} {1983})}\BibitemShut
  {NoStop}%
\bibitem [{\citenamefont {Weiss}(2008)}]{Wei08}%
  \BibitemOpen
  \bibfield  {author} {\bibinfo {author} {\bibfnamefont {U.}~\bibnamefont
  {Weiss}},\ }\href@noop {} {\emph {\bibinfo {title} {Quantum Dissipative
  Systems}}},\ \bibinfo {edition} {3rd}\ ed.\ (\bibinfo  {publisher} {World
  Scientific, Singapore},\ \bibinfo {year} {2008})\BibitemShut {NoStop}%
\bibitem [{\citenamefont {Gardiner}\ and\ \citenamefont {Zoller}(2010)}]{GZ10}%
  \BibitemOpen
  \bibfield  {author} {\bibinfo {author} {\bibfnamefont {C.~W.}\ \bibnamefont
  {Gardiner}}\ and\ \bibinfo {author} {\bibfnamefont {P.}~\bibnamefont
  {Zoller}},\ }\href@noop {} {\emph {\bibinfo {title} {Quantum Noise: A
  Handbook of Markovian and non-Markovian Quantum Stochastic Methods}}},\
  \bibinfo {edition} {3rd}\ ed.\ (\bibinfo  {publisher} {Springer, Berlin
  Heidelberg},\ \bibinfo {year} {2010})\BibitemShut {NoStop}%
\bibitem [{\citenamefont {Breuer}\ and\ \citenamefont
  {Petruccione}(2002)}]{BP02}%
  \BibitemOpen
  \bibfield  {author} {\bibinfo {author} {\bibfnamefont {H.-P.}\ \bibnamefont
  {Breuer}}\ and\ \bibinfo {author} {\bibfnamefont {F.}~\bibnamefont
  {Petruccione}},\ }\href@noop {} {\emph {\bibinfo {title} {The Theory of Open
  Quantum Systems}}}\ (\bibinfo  {publisher} {Oxford University Press},\
  \bibinfo {year} {2002})\BibitemShut {NoStop}%
\bibitem [{\citenamefont {Mukamel}(1999)}]{Muk99}%
  \BibitemOpen
  \bibfield  {author} {\bibinfo {author} {\bibfnamefont {S.}~\bibnamefont
  {Mukamel}},\ }\href@noop {} {\emph {\bibinfo {title} {Principles of Nonlinear
  Optical Spectroscopy}}}\ (\bibinfo  {publisher} {Oxford University Press},\
  \bibinfo {address} {New York},\ \bibinfo {year} {1999})\BibitemShut {NoStop}%
\bibitem [{\citenamefont {May}\ and\ \citenamefont {K\"uhn}(2011)}]{MK11}%
  \BibitemOpen
  \bibfield  {author} {\bibinfo {author} {\bibfnamefont {V.}~\bibnamefont
  {May}}\ and\ \bibinfo {author} {\bibfnamefont {O.}~\bibnamefont {K\"uhn}},\
  }\href@noop {} {\emph {\bibinfo {title} {Charge and energy transfer dynamics
  in molecular systems}}},\ \bibinfo {edition} {3rd}\ ed.\ (\bibinfo
  {publisher} {Wiley-VCH, Weinheim},\ \bibinfo {year} {2011})\BibitemShut
  {NoStop}%
\bibitem [{\citenamefont {Zurek}(2003)}]{Zur03}%
  \BibitemOpen
  \bibfield  {author} {\bibinfo {author} {\bibfnamefont {W.~H.}\ \bibnamefont
  {Zurek}},\ }\href {\doibase 10.1103/RevModPhys.75.715} {\bibfield  {journal}
  {\bibinfo  {journal} {Rev. Mod. Phys.}\ }\textbf {\bibinfo {volume} {75}},\
  \bibinfo {pages} {715} (\bibinfo {year} {2003})}\BibitemShut {NoStop}%
\bibitem [{\citenamefont {Erez}\ \emph {et~al.}(2008)\citenamefont {Erez},
  \citenamefont {Gordon}, \citenamefont {Nest},\ and\ \citenamefont
  {Kurizki}}]{EG&08}%
  \BibitemOpen
  \bibfield  {author} {\bibinfo {author} {\bibfnamefont {N.}~\bibnamefont
  {Erez}}, \bibinfo {author} {\bibfnamefont {G.}~\bibnamefont {Gordon}},
  \bibinfo {author} {\bibfnamefont {M.}~\bibnamefont {Nest}}, \ and\ \bibinfo
  {author} {\bibfnamefont {G.}~\bibnamefont {Kurizki}},\ }\href {\doibase
  10.1038/nature06873} {\bibfield  {journal} {\bibinfo  {journal} {Nature}\
  }\textbf {\bibinfo {volume} {452}},\ \bibinfo {pages} {724} (\bibinfo {year}
  {2008})}\BibitemShut {NoStop}%
\bibitem [{\citenamefont {Galve}\ \emph {et~al.}(2010)\citenamefont {Galve},
  \citenamefont {Pach\'on},\ and\ \citenamefont {Zueco}}]{GPZ10}%
  \BibitemOpen
  \bibfield  {author} {\bibinfo {author} {\bibfnamefont {F.}~\bibnamefont
  {Galve}}, \bibinfo {author} {\bibfnamefont {L.~A.}\ \bibnamefont {Pach\'on}},
  \ and\ \bibinfo {author} {\bibfnamefont {D.}~\bibnamefont {Zueco}},\ }\href
  {\doibase 10.1103/PhysRevLett.105.180501} {\bibfield  {journal} {\bibinfo
  {journal} {Phys. Rev. Lett.}\ }\textbf {\bibinfo {volume} {105}},\ \bibinfo
  {pages} {180501} (\bibinfo {year} {2010})},\ \Eprint
  {http://arxiv.org/abs/1002.1923} {arXiv:1002.1923} \BibitemShut {NoStop}%
\bibitem [{\citenamefont {Pach\'on}\ and\ \citenamefont {Brumer}(2011)}]{PB11}%
  \BibitemOpen
  \bibfield  {author} {\bibinfo {author} {\bibfnamefont {L.~A.}\ \bibnamefont
  {Pach\'on}}\ and\ \bibinfo {author} {\bibfnamefont {P.}~\bibnamefont
  {Brumer}},\ }\href {\doibase 10.1021/jz201189p} {\bibfield  {journal}
  {\bibinfo  {journal} {J. Phys. Chem. Lett.}\ }\textbf {\bibinfo {volume}
  {2}},\ \bibinfo {pages} {2728} (\bibinfo {year} {2011})}\BibitemShut
  {NoStop}%
\bibitem [{\citenamefont {Grabert}\ \emph {et~al.}(1988)\citenamefont
  {Grabert}, \citenamefont {Schramm},\ and\ \citenamefont {Ingold}}]{GSI88}%
  \BibitemOpen
  \bibfield  {author} {\bibinfo {author} {\bibfnamefont {H.}~\bibnamefont
  {Grabert}}, \bibinfo {author} {\bibfnamefont {P.}~\bibnamefont {Schramm}}, \
  and\ \bibinfo {author} {\bibfnamefont {G.-L.}\ \bibnamefont {Ingold}},\
  }\href {\doibase 10.1016/0370-1573(88)90023-3} {\bibfield  {journal}
  {\bibinfo  {journal} {Phys. Rep.}\ }\textbf {\bibinfo {volume} {168}},\
  \bibinfo {pages} {115} (\bibinfo {year} {1988})}\BibitemShut {NoStop}%
\bibitem [{\citenamefont {Pach\'on}\ and\ \citenamefont
  {Brumer}(2012)}]{PB12c}%
  \BibitemOpen
  \bibfield  {author} {\bibinfo {author} {\bibfnamefont {L.~A.}\ \bibnamefont
  {Pach\'on}}\ and\ \bibinfo {author} {\bibfnamefont {P.}~\bibnamefont
  {Brumer}},\ }\href {\doibase 10.1039/c2cp40815e} {\bibfield  {journal}
  {\bibinfo  {journal} {Phys. Chem. Chem. Phys.}\ }\textbf {\bibinfo {volume}
  {14}},\ \bibinfo {pages} {10094} (\bibinfo {year} {2012})},\ \Eprint
  {http://arxiv.org/abs/1203.3978} {arXiv:1203.3978} \BibitemShut {NoStop}%
\bibitem [{\citenamefont {Singh}\ and\ \citenamefont {Brumer}(2012)}]{SB12}%
  \BibitemOpen
  \bibfield  {author} {\bibinfo {author} {\bibfnamefont {N.}~\bibnamefont
  {Singh}}\ and\ \bibinfo {author} {\bibfnamefont {P.}~\bibnamefont {Brumer}},\
  }\href@noop {} {\bibfield  {journal} {\bibinfo  {journal} {Mol. Phys.}\
  }\textbf {\bibinfo {volume} {110}},\ \bibinfo {pages} {1815} (\bibinfo {year}
  {2012})}\BibitemShut {NoStop}%
\bibitem [{\citenamefont {H\"{a}nggi}\ and\ \citenamefont
  {Ingold}(2005)}]{HI05}%
  \BibitemOpen
  \bibfield  {author} {\bibinfo {author} {\bibfnamefont {P.}~\bibnamefont
  {H\"{a}nggi}}\ and\ \bibinfo {author} {\bibfnamefont {G.-L.}\ \bibnamefont
  {Ingold}},\ }\href {\doibase 10.1063/1.1853631} {\bibfield  {journal}
  {\bibinfo  {journal} {Chaos}\ }\textbf {\bibinfo {volume} {15}},\ \bibinfo
  {pages} {026105} (\bibinfo {year} {2005})}\BibitemShut {NoStop}%
\bibitem [{\citenamefont {Zerbe}\ and\ \citenamefont {H\"anggi}(1995)}]{ZH95}%
  \BibitemOpen
  \bibfield  {author} {\bibinfo {author} {\bibfnamefont {C.}~\bibnamefont
  {Zerbe}}\ and\ \bibinfo {author} {\bibfnamefont {P.}~\bibnamefont
  {H\"anggi}},\ }\href@noop {} {\bibfield  {journal} {\bibinfo  {journal}
  {Phys. Rev. E}\ }\textbf {\bibinfo {volume} {52}},\ \bibinfo {pages} {1533}
  (\bibinfo {year} {1995})}\BibitemShut {NoStop}%
\bibitem [{\citenamefont {Kofman}\ and\ \citenamefont {Kurizki}(2004)}]{KK04}%
  \BibitemOpen
  \bibfield  {author} {\bibinfo {author} {\bibfnamefont {A.~G.}\ \bibnamefont
  {Kofman}}\ and\ \bibinfo {author} {\bibfnamefont {G.}~\bibnamefont
  {Kurizki}},\ }\href {\doibase 10.1103/PhysRevLett.93.130406} {\bibfield
  {journal} {\bibinfo  {journal} {Phys. Rev. Lett.}\ }\textbf {\bibinfo
  {volume} {93}},\ \bibinfo {pages} {130406} (\bibinfo {year}
  {2004})}\BibitemShut {NoStop}%
\bibitem [{\citenamefont {Schmidt}\ \emph {et~al.}(2011)\citenamefont
  {Schmidt}, \citenamefont {Negretti}, \citenamefont {Ankerhold}, \citenamefont
  {Calarco},\ and\ \citenamefont {Stockburger}}]{SN&11}%
  \BibitemOpen
  \bibfield  {author} {\bibinfo {author} {\bibfnamefont {R.}~\bibnamefont
  {Schmidt}}, \bibinfo {author} {\bibfnamefont {A.}~\bibnamefont {Negretti}},
  \bibinfo {author} {\bibfnamefont {J.}~\bibnamefont {Ankerhold}}, \bibinfo
  {author} {\bibfnamefont {T.}~\bibnamefont {Calarco}}, \ and\ \bibinfo
  {author} {\bibfnamefont {J.~T.}\ \bibnamefont {Stockburger}},\ }\href
  {\doibase 10.1103/PhysRevLett.107.130404} {\bibfield  {journal} {\bibinfo
  {journal} {Phys. Rev. Lett.}\ }\textbf {\bibinfo {volume} {107}},\ \bibinfo
  {pages} {130404} (\bibinfo {year} {2011})}\BibitemShut {NoStop}%
\bibitem [{\citenamefont {Shapiro}\ and\ \citenamefont {Brumer}(2011)}]{SB11}%
  \BibitemOpen
  \bibfield  {author} {\bibinfo {author} {\bibfnamefont {M.}~\bibnamefont
  {Shapiro}}\ and\ \bibinfo {author} {\bibfnamefont {P.}~\bibnamefont
  {Brumer}},\ }\href@noop {} {\emph {\bibinfo {title} {Quantum Control of
  Molecular Processes}}},\ \bibinfo {edition} {2nd}\ ed.\ (\bibinfo
  {publisher} {Wiley-VCH},\ \bibinfo {year} {2011})\BibitemShut {NoStop}%
\bibitem [{\citenamefont {Pach\'on}\ and\ \citenamefont
  {Brumer}(2013{\natexlab{a}})}]{PB13}%
  \BibitemOpen
  \bibfield  {author} {\bibinfo {author} {\bibfnamefont {L.~A.}\ \bibnamefont
  {Pach\'on}}\ and\ \bibinfo {author} {\bibfnamefont {P.}~\bibnamefont
  {Brumer}},\ }\href {\doibase 10.1103/PhysRevA.87.022106} {\bibfield
  {journal} {\bibinfo  {journal} {Phys. Rev. A}\ }\textbf {\bibinfo {volume}
  {87}},\ \bibinfo {pages} {022106} (\bibinfo {year} {2013}{\natexlab{a}})},\
  \Eprint {http://arxiv.org/abs/1210.6374} {arXiv:1210.6374} \BibitemShut
  {NoStop}%
\bibitem [{\citenamefont {Feynman}\ and\ \citenamefont {Hibbs}(1965)}]{FH65}%
  \BibitemOpen
  \bibfield  {author} {\bibinfo {author} {\bibfnamefont {R.~P.}\ \bibnamefont
  {Feynman}}\ and\ \bibinfo {author} {\bibfnamefont {A.~R.}\ \bibnamefont
  {Hibbs}},\ }\href@noop {} {\emph {\bibinfo {title} {Quantum physics and path
  integrals}}}\ (\bibinfo  {publisher} {McGraw--Hill, New York},\ \bibinfo
  {year} {1965})\BibitemShut {NoStop}%
\bibitem [{\citenamefont {Jiang}\ and\ \citenamefont {Brumer}(1991)}]{JB91}%
  \BibitemOpen
  \bibfield  {author} {\bibinfo {author} {\bibfnamefont {X.-P.}\ \bibnamefont
  {Jiang}}\ and\ \bibinfo {author} {\bibfnamefont {P.}~\bibnamefont {Brumer}},\
  }\href {\doibase 10.1063/1.460467} {\bibfield  {journal} {\bibinfo  {journal}
  {J. Chem. Phys.}\ }\textbf {\bibinfo {volume} {94}},\ \bibinfo {pages} {5833}
  (\bibinfo {year} {1991})}\BibitemShut {NoStop}%
\bibitem [{\citenamefont {Brumer}\ and\ \citenamefont {Shapiro}(2012)}]{BS11b}%
  \BibitemOpen
  \bibfield  {author} {\bibinfo {author} {\bibfnamefont {P.}~\bibnamefont
  {Brumer}}\ and\ \bibinfo {author} {\bibfnamefont {M.}~\bibnamefont
  {Shapiro}},\ }\href {\doibase 10.1073/pnas.1211209109} {\bibfield  {journal}
  {\bibinfo  {journal} {Proc. Natl. Acad. Sci. U.S.A.}\ }\textbf {\bibinfo
  {volume} {109}},\ \bibinfo {pages} {19575} (\bibinfo {year}
  {2012})}\BibitemShut {NoStop}%
\bibitem [{\citenamefont {Engel}\ \emph {et~al.}(2007)\citenamefont {Engel}
  \emph {et~al.}}]{EC&07}%
  \BibitemOpen
  \bibfield  {author} {\bibinfo {author} {\bibfnamefont {G.~S.}\ \bibnamefont
  {Engel}} \emph {et~al.},\ }\href {\doibase 10.1038/nature05678} {\bibfield
  {journal} {\bibinfo  {journal} {Nature}\ }\textbf {\bibinfo {volume} {446}},\
  \bibinfo {pages} {782} (\bibinfo {year} {2007})}\BibitemShut {NoStop}%
\bibitem [{\citenamefont {Hoki}\ and\ \citenamefont {Brumer}(2011)}]{HB11}%
  \BibitemOpen
  \bibfield  {author} {\bibinfo {author} {\bibfnamefont {K.}~\bibnamefont
  {Hoki}}\ and\ \bibinfo {author} {\bibfnamefont {P.}~\bibnamefont {Brumer}},\
  }\href@noop {} {\bibfield  {journal} {\bibinfo  {journal} {Proc. Chem.}\
  }\textbf {\bibinfo {volume} {3}},\ \bibinfo {pages} {122} (\bibinfo {year}
  {2011})}\BibitemShut {NoStop}%
\bibitem [{\citenamefont {Spanner}\ \emph {et~al.}(2010)\citenamefont
  {Spanner}, \citenamefont {Arango},\ and\ \citenamefont {Brumer}}]{SAB10}%
  \BibitemOpen
  \bibfield  {author} {\bibinfo {author} {\bibfnamefont {M.}~\bibnamefont
  {Spanner}}, \bibinfo {author} {\bibfnamefont {C.~A.}\ \bibnamefont {Arango}},
  \ and\ \bibinfo {author} {\bibfnamefont {P.}~\bibnamefont {Brumer}},\ }\href
  {\doibase 10.1063/1.3491366} {\bibfield  {journal} {\bibinfo  {journal} {J.
  Chem. Phys.}\ }\textbf {\bibinfo {volume} {133}},\ \bibinfo {pages} {151101}
  (\bibinfo {year} {2010})}\BibitemShut {NoStop}%
\bibitem [{\citenamefont {Pach\'on}\ \emph {et~al.}(2013)\citenamefont
  {Pach\'on}, \citenamefont {Yu},\ and\ \citenamefont {Brumer}}]{PYB13}%
  \BibitemOpen
  \bibfield  {author} {\bibinfo {author} {\bibfnamefont {L.~A.}\ \bibnamefont
  {Pach\'on}}, \bibinfo {author} {\bibfnamefont {L.}~\bibnamefont {Yu}}, \ and\
  \bibinfo {author} {\bibfnamefont {P.}~\bibnamefont {Brumer}},\ }\href
  {\doibase 10.1039/C3FD20144A} {\bibfield  {journal} {\bibinfo  {journal}
  {Faraday Discussions}\ }\textbf {\bibinfo {volume} {163}},\ \bibinfo {pages}
  {485} (\bibinfo {year} {2013})},\ \Eprint {http://arxiv.org/abs/1212.6416}
  {arXiv:1212.6416} \BibitemShut {NoStop}%
\bibitem [{\citenamefont {Pach\'on}\ and\ \citenamefont
  {Brumer}(2013{\natexlab{b}})}]{PB13c}%
  \BibitemOpen
  \bibfield  {author} {\bibinfo {author} {\bibfnamefont {L.~A.}\ \bibnamefont
  {Pach\'on}}\ and\ \bibinfo {author} {\bibfnamefont {P.}~\bibnamefont
  {Brumer}},\ }\href {\doibase http://dx.doi.org/10.1063/1.4825358} {\bibfield
  {journal} {\bibinfo  {journal} {J. Chem. Phys.}\ }\textbf {\bibinfo {volume}
  {139}},\ \bibinfo {eid} {164123} (\bibinfo {year} {2013}{\natexlab{b}})},\
  \Eprint {http://arxiv.org/abs/1308.1843} {arXiv:1308.1843} \BibitemShut
  {NoStop}%
\bibitem [{\citenamefont {Ford}\ \emph {et~al.}(1985)\citenamefont {Ford},
  \citenamefont {Lewis},\ and\ \citenamefont {O'Connell}}]{FLO85}%
  \BibitemOpen
  \bibfield  {author} {\bibinfo {author} {\bibfnamefont {G.}~\bibnamefont
  {Ford}}, \bibinfo {author} {\bibfnamefont {J.}~\bibnamefont {Lewis}}, \ and\
  \bibinfo {author} {\bibfnamefont {R.}~\bibnamefont {O'Connell}},\ }\href
  {\doibase 10.1103/PhysRevLett.55.2273} {\bibfield  {journal} {\bibinfo
  {journal} {Phys. Rev. Lett.}\ }\textbf {\bibinfo {volume} {55}},\ \bibinfo
  {pages} {2273} (\bibinfo {year} {1985})}\BibitemShut {NoStop}%
\bibitem [{\citenamefont {Ford}\ \emph
  {et~al.}(1988{\natexlab{a}})\citenamefont {Ford}, \citenamefont {Lewis},\
  and\ \citenamefont {O'Connell}}]{FLO88}%
  \BibitemOpen
  \bibfield  {author} {\bibinfo {author} {\bibfnamefont {G.}~\bibnamefont
  {Ford}}, \bibinfo {author} {\bibfnamefont {J.}~\bibnamefont {Lewis}}, \ and\
  \bibinfo {author} {\bibfnamefont {R.}~\bibnamefont {O'Connell}},\ }\href
  {\doibase 10.1103/PhysRevA.37.4419} {\bibfield  {journal} {\bibinfo
  {journal} {Phys. Rev. A}\ }\textbf {\bibinfo {volume} {37}},\ \bibinfo
  {pages} {4419} (\bibinfo {year} {1988}{\natexlab{a}})}\BibitemShut {NoStop}%
\bibitem [{\citenamefont {Castrigiano}\ and\ \citenamefont
  {Kokiantonis}(1987)}]{CK87}%
  \BibitemOpen
  \bibfield  {author} {\bibinfo {author} {\bibfnamefont {D.~P.~L.}\
  \bibnamefont {Castrigiano}}\ and\ \bibinfo {author} {\bibfnamefont
  {N.}~\bibnamefont {Kokiantonis}},\ }\href {\doibase 10.1103/PhysRevA.35.4122}
  {\bibfield  {journal} {\bibinfo  {journal} {Phys. Rev. A}\ }\textbf {\bibinfo
  {volume} {35}},\ \bibinfo {pages} {4122} (\bibinfo {year}
  {1987})}\BibitemShut {NoStop}%
\bibitem [{\citenamefont {Barone}\ and\ \citenamefont {Caldeira}(1991)}]{BC91}%
  \BibitemOpen
  \bibfield  {author} {\bibinfo {author} {\bibfnamefont {P.~M. V.~B.}\
  \bibnamefont {Barone}}\ and\ \bibinfo {author} {\bibfnamefont {A.~O.}\
  \bibnamefont {Caldeira}},\ }\href {\doibase 10.1103/PhysRevA.43.57}
  {\bibfield  {journal} {\bibinfo  {journal} {Phys. Rev. A}\ }\textbf {\bibinfo
  {volume} {43}},\ \bibinfo {pages} {57} (\bibinfo {year} {1991})}\BibitemShut
  {NoStop}%
\bibitem [{\citenamefont {Ford}\ \emph
  {et~al.}(1988{\natexlab{b}})\citenamefont {Ford}, \citenamefont {Lewis},\
  and\ \citenamefont {O'Connell}}]{FLO88b}%
  \BibitemOpen
  \bibfield  {author} {\bibinfo {author} {\bibfnamefont {G.~W.}\ \bibnamefont
  {Ford}}, \bibinfo {author} {\bibfnamefont {J.~T.}\ \bibnamefont {Lewis}}, \
  and\ \bibinfo {author} {\bibfnamefont {R.~F.}\ \bibnamefont {O'Connell}},\
  }\href {\doibase 10.1103/PhysRevA.37.3609} {\bibfield  {journal} {\bibinfo
  {journal} {Phys. Rev. A}\ }\textbf {\bibinfo {volume} {37}},\ \bibinfo
  {pages} {3609} (\bibinfo {year} {1988}{\natexlab{b}})}\BibitemShut {NoStop}%
\bibitem [{\citenamefont {Castrigiano}\ and\ \citenamefont
  {Kokiantonis}(1988)}]{CK88}%
  \BibitemOpen
  \bibfield  {author} {\bibinfo {author} {\bibfnamefont {D.~P.~L.}\
  \bibnamefont {Castrigiano}}\ and\ \bibinfo {author} {\bibfnamefont
  {N.}~\bibnamefont {Kokiantonis}},\ }\href {\doibase 10.1103/PhysRevA.38.527}
  {\bibfield  {journal} {\bibinfo  {journal} {Phys. Rev. A}\ }\textbf {\bibinfo
  {volume} {38}},\ \bibinfo {pages} {527} (\bibinfo {year} {1988})}\BibitemShut
  {NoStop}%
\bibitem [{\citenamefont {Power}\ and\ \citenamefont {Zienau}(1959)}]{PZ59}%
  \BibitemOpen
  \bibfield  {author} {\bibinfo {author} {\bibfnamefont {E.~A.}\ \bibnamefont
  {Power}}\ and\ \bibinfo {author} {\bibfnamefont {S.}~\bibnamefont {Zienau}},\
  }\href {\doibase 10.1098/rsta.1959.0008} {\bibfield  {journal} {\bibinfo
  {journal} {Phil. Trans. R. Soc. Lond. A}\ }\textbf {\bibinfo {volume}
  {251}},\ \bibinfo {pages} {427} (\bibinfo {year} {1959})}\BibitemShut
  {NoStop}%
\bibitem [{\citenamefont {Herzfeld}\ and\ \citenamefont
  {Goeppert-Mayer}(1936)}]{HG36}%
  \BibitemOpen
  \bibfield  {author} {\bibinfo {author} {\bibfnamefont {K.~F.}\ \bibnamefont
  {Herzfeld}}\ and\ \bibinfo {author} {\bibfnamefont {M.}~\bibnamefont
  {Goeppert-Mayer}},\ }\href {\doibase 10.1103/PhysRev.49.332} {\bibfield
  {journal} {\bibinfo  {journal} {Phys. Rev.}\ }\textbf {\bibinfo {volume}
  {49}},\ \bibinfo {pages} {332} (\bibinfo {year} {1936})}\BibitemShut
  {NoStop}%
\bibitem [{\citenamefont {Yang}\ \emph {et~al.}(1981)\citenamefont {Yang},
  \citenamefont {Hirschfelder},\ and\ \citenamefont {Johnson}}]{YHJ81}%
  \BibitemOpen
  \bibfield  {author} {\bibinfo {author} {\bibfnamefont {K.-H.~T.}\
  \bibnamefont {Yang}}, \bibinfo {author} {\bibfnamefont {J.~O.}\ \bibnamefont
  {Hirschfelder}}, \ and\ \bibinfo {author} {\bibfnamefont {B.~R.}\
  \bibnamefont {Johnson}},\ }\href {\doibase 10.1063/1.442295} {\bibfield
  {journal} {\bibinfo  {journal} {J. Chem. Phys.}\ }\textbf {\bibinfo {volume}
  {75}},\ \bibinfo {pages} {2321} (\bibinfo {year} {1981})}\BibitemShut
  {NoStop}%
\bibitem [{\citenamefont {Ingold}(2002)}]{Ing02}%
  \BibitemOpen
  \bibfield  {author} {\bibinfo {author} {\bibfnamefont {G.-L.}\ \bibnamefont
  {Ingold}},\ }in\ \href@noop {} {\emph {\bibinfo {booktitle} {Coherent
  Evolution in Noisy Environments}}},\ \bibinfo {series} {Lecture Notes in
  Physics}, Vol.\ \bibinfo {volume} {611},\ \bibinfo {editor} {edited by\
  \bibinfo {editor} {\bibfnamefont {A.}~\bibnamefont {Buchleitner}}\ and\
  \bibinfo {editor} {\bibfnamefont {K.}~\bibnamefont {Hornberger}}}\ (\bibinfo
  {publisher} {Springer Verlag},\ \bibinfo {address} {Berlin-Heidelberg-New
  York},\ \bibinfo {year} {2002})\BibitemShut {NoStop}%
\bibitem [{\citenamefont {Schramm}\ and\ \citenamefont {Grabert}(1986)}]{SG86}%
  \BibitemOpen
  \bibfield  {author} {\bibinfo {author} {\bibfnamefont {P.}~\bibnamefont
  {Schramm}}\ and\ \bibinfo {author} {\bibfnamefont {H.}~\bibnamefont
  {Grabert}},\ }\href {\doibase 10.1103/PhysRevA.34.4515} {\bibfield  {journal}
  {\bibinfo  {journal} {Phy. Rev. A}\ }\textbf {\bibinfo {volume} {34}},\
  \bibinfo {pages} {4515} (\bibinfo {year} {1986})}\BibitemShut {NoStop}%
\bibitem [{\citenamefont {Pach\'on}(2010)}]{Pac10}%
  \BibitemOpen
  \bibfield  {author} {\bibinfo {author} {\bibfnamefont {L.~A.}\ \bibnamefont
  {Pach\'on}},\ }\emph {\bibinfo {title} {Coherence and Decoherence in the
  Semiclassical propagation of the Wigner function}},\ \href@noop {} {Ph.D.
  thesis},\ \bibinfo  {school} {Universidad Nacional de Colombia} (\bibinfo
  {year} {2010})\BibitemShut {NoStop}%
\bibitem [{\citenamefont {Ford}\ \emph {et~al.}(1987)\citenamefont {Ford},
  \citenamefont {Lewis},\ and\ \citenamefont {O'Connell}}]{FLO87}%
  \BibitemOpen
  \bibfield  {author} {\bibinfo {author} {\bibfnamefont {G.}~\bibnamefont
  {Ford}}, \bibinfo {author} {\bibfnamefont {J.}~\bibnamefont {Lewis}}, \ and\
  \bibinfo {author} {\bibfnamefont {R.}~\bibnamefont {O'Connell}},\ }\href
  {\doibase 10.1103/PhysRevA.36.1466} {\bibfield  {journal} {\bibinfo
  {journal} {Phys. Rev. A}\ }\textbf {\bibinfo {volume} {36}},\ \bibinfo
  {pages} {1466} (\bibinfo {year} {1987})}\BibitemShut {NoStop}%
\bibitem [{\citenamefont {Mehta}\ and\ \citenamefont
  {Wolf}(1964{\natexlab{a}})}]{MW64a}%
  \BibitemOpen
  \bibfield  {author} {\bibinfo {author} {\bibfnamefont {C.~L.}\ \bibnamefont
  {Mehta}}\ and\ \bibinfo {author} {\bibfnamefont {E.}~\bibnamefont {Wolf}},\
  }\href {\doibase 10.1103/PhysRev.134.A1143} {\bibfield  {journal} {\bibinfo
  {journal} {Phys. Rev.}\ }\textbf {\bibinfo {volume} {134}},\ \bibinfo {pages}
  {A1143} (\bibinfo {year} {1964}{\natexlab{a}})}\BibitemShut {NoStop}%
\bibitem [{\citenamefont {Mehta}\ and\ \citenamefont
  {Wolf}(1964{\natexlab{b}})}]{MW64b}%
  \BibitemOpen
  \bibfield  {author} {\bibinfo {author} {\bibfnamefont {C.~L.}\ \bibnamefont
  {Mehta}}\ and\ \bibinfo {author} {\bibfnamefont {E.}~\bibnamefont {Wolf}},\
  }\href {\doibase 10.1103/PhysRev.134.A1149} {\bibfield  {journal} {\bibinfo
  {journal} {Phys. Rev.}\ }\textbf {\bibinfo {volume} {134}},\ \bibinfo {pages}
  {A1149} (\bibinfo {year} {1964}{\natexlab{b}})}\BibitemShut {NoStop}%
\bibitem [{\citenamefont {Mehta}\ and\ \citenamefont {Wolf}(1967)}]{MW67}%
  \BibitemOpen
  \bibfield  {author} {\bibinfo {author} {\bibfnamefont {C.~L.}\ \bibnamefont
  {Mehta}}\ and\ \bibinfo {author} {\bibfnamefont {E.}~\bibnamefont {Wolf}},\
  }\href {\doibase 10.1103/PhysRev.161.1328} {\bibfield  {journal} {\bibinfo
  {journal} {Phys. Rev.}\ }\textbf {\bibinfo {volume} {161}},\ \bibinfo {pages}
  {1328} (\bibinfo {year} {1967})}\BibitemShut {NoStop}%
\bibitem [{\citenamefont {Mandel}\ and\ \citenamefont {Wolf}(1995)}]{MW95}%
  \BibitemOpen
  \bibfield  {author} {\bibinfo {author} {\bibfnamefont {L.}~\bibnamefont
  {Mandel}}\ and\ \bibinfo {author} {\bibfnamefont {E.}~\bibnamefont {Wolf}},\
  }\href@noop {} {\emph {\bibinfo {title} {Optical coherence and quantum
  optics}}}\ (\bibinfo  {publisher} {Cambridge University Press, Cambridge},\
  \bibinfo {year} {1995})\BibitemShut {NoStop}%
\bibitem [{\citenamefont {Ford}\ and\ \citenamefont {O'Connell}(1991)}]{FO91}%
  \BibitemOpen
  \bibfield  {author} {\bibinfo {author} {\bibfnamefont {G.}~\bibnamefont
  {Ford}}\ and\ \bibinfo {author} {\bibfnamefont {R.}~\bibnamefont
  {O'Connell}},\ }\href {\doibase 10.1016/0375-9601(91)90054-C} {\bibfield
  {journal} {\bibinfo  {journal} {Phys. Lett. A}\ }\textbf {\bibinfo {volume}
  {45}},\ \bibinfo {pages} {217} (\bibinfo {year} {1991})}\BibitemShut
  {NoStop}%
\bibitem [{\citenamefont {Ford}\ and\ \citenamefont {O'Connell}(1998)}]{FO98}%
  \BibitemOpen
  \bibfield  {author} {\bibinfo {author} {\bibfnamefont {G.~W.}\ \bibnamefont
  {Ford}}\ and\ \bibinfo {author} {\bibfnamefont {R.~F.}\ \bibnamefont
  {O'Connell}},\ }\href {\doibase 10.1103/PhysRevA.57.3112} {\bibfield
  {journal} {\bibinfo  {journal} {Phys. Rev. A}\ }\textbf {\bibinfo {volume}
  {57}},\ \bibinfo {pages} {3112} (\bibinfo {year} {1998})}\BibitemShut
  {NoStop}%
\bibitem [{\citenamefont {Hakim}\ and\ \citenamefont
  {Ambegaokar}(1985)}]{HA85}%
  \BibitemOpen
  \bibfield  {author} {\bibinfo {author} {\bibfnamefont {V.}~\bibnamefont
  {Hakim}}\ and\ \bibinfo {author} {\bibfnamefont {V.}~\bibnamefont
  {Ambegaokar}},\ }\href {\doibase 10.1103/PhysRevA.32.423} {\bibfield
  {journal} {\bibinfo  {journal} {Phys. Rev. A}\ }\textbf {\bibinfo {volume}
  {32}},\ \bibinfo {pages} {423} (\bibinfo {year} {1985})}\BibitemShut
  {NoStop}%
\bibitem [{\citenamefont {Pach\'on}\ \emph {et~al.}(2010)\citenamefont
  {Pach\'on}, \citenamefont {Ingold},\ and\ \citenamefont {Dittrich}}]{PID10}%
  \BibitemOpen
  \bibfield  {author} {\bibinfo {author} {\bibfnamefont {L.~A.}\ \bibnamefont
  {Pach\'on}}, \bibinfo {author} {\bibfnamefont {G.-L.}\ \bibnamefont
  {Ingold}}, \ and\ \bibinfo {author} {\bibfnamefont {T.}~\bibnamefont
  {Dittrich}},\ }\href {\doibase 10.1016/j.chemphys.2010.05.024} {\bibfield
  {journal} {\bibinfo  {journal} {Chem. Phys.}\ }\textbf {\bibinfo {volume}
  {375}},\ \bibinfo {pages} {209} (\bibinfo {year} {2010})},\ \Eprint
  {http://arxiv.org/abs/1005.3839} {arXiv:1005.3839} \BibitemShut {NoStop}%
\bibitem [{\citenamefont {Li}\ \emph {et~al.}(1990{\natexlab{a}})\citenamefont
  {Li}, \citenamefont {Ford},\ and\ \citenamefont {O'Connell}}]{LFO90}%
  \BibitemOpen
  \bibfield  {author} {\bibinfo {author} {\bibfnamefont {X.~L.}\ \bibnamefont
  {Li}}, \bibinfo {author} {\bibfnamefont {G.~W.}\ \bibnamefont {Ford}}, \ and\
  \bibinfo {author} {\bibfnamefont {R.~F.}\ \bibnamefont {O'Connell}},\ }\href
  {\doibase 10.1103/PhysRevA.41.5287} {\bibfield  {journal} {\bibinfo
  {journal} {Phys. Rev. A}\ }\textbf {\bibinfo {volume} {41}},\ \bibinfo
  {pages} {5287} (\bibinfo {year} {1990}{\natexlab{a}})}\BibitemShut {NoStop}%
\bibitem [{\citenamefont {Li}\ \emph {et~al.}(1990{\natexlab{b}})\citenamefont
  {Li}, \citenamefont {Ford},\ and\ \citenamefont {O'Connell}}]{LFO90b}%
  \BibitemOpen
  \bibfield  {author} {\bibinfo {author} {\bibfnamefont {X.~L.}\ \bibnamefont
  {Li}}, \bibinfo {author} {\bibfnamefont {G.~W.}\ \bibnamefont {Ford}}, \ and\
  \bibinfo {author} {\bibfnamefont {R.~F.}\ \bibnamefont {O'Connell}},\ }\href
  {\doibase 10.1103/PhysRevA.42.4519} {\bibfield  {journal} {\bibinfo
  {journal} {Phys. Rev. A}\ }\textbf {\bibinfo {volume} {42}},\ \bibinfo
  {pages} {4519} (\bibinfo {year} {1990}{\natexlab{b}})}\BibitemShut {NoStop}%
\bibitem [{\citenamefont {Grabert}\ \emph {et~al.}(1984)\citenamefont
  {Grabert}, \citenamefont {Weiss},\ and\ \citenamefont {Talkner}}]{GWT84}%
  \BibitemOpen
  \bibfield  {author} {\bibinfo {author} {\bibfnamefont {H.}~\bibnamefont
  {Grabert}}, \bibinfo {author} {\bibfnamefont {U.}~\bibnamefont {Weiss}}, \
  and\ \bibinfo {author} {\bibfnamefont {P.}~\bibnamefont {Talkner}},\ }\href
  {\doibase 10.1007/BF01307505} {\bibfield  {journal} {\bibinfo  {journal} {Z.
  Phys. B}\ }\textbf {\bibinfo {volume} {55}},\ \bibinfo {pages} {87} (\bibinfo
  {year} {1984})}\BibitemShut {NoStop}%
\bibitem [{\citenamefont {Hu}\ \emph {et~al.}(1992)\citenamefont {Hu},
  \citenamefont {Paz},\ and\ \citenamefont {Zhang}}]{HPZ92}%
  \BibitemOpen
  \bibfield  {author} {\bibinfo {author} {\bibfnamefont {B.~L.}\ \bibnamefont
  {Hu}}, \bibinfo {author} {\bibfnamefont {J.~P.}\ \bibnamefont {Paz}}, \ and\
  \bibinfo {author} {\bibfnamefont {Y.}~\bibnamefont {Zhang}},\ }\href
  {\doibase 10.1103/PhysRevD.45.2843} {\bibfield  {journal} {\bibinfo
  {journal} {Phys. Rev. D}\ }\textbf {\bibinfo {volume} {45}},\ \bibinfo
  {pages} {2843} (\bibinfo {year} {1992})}\BibitemShut {NoStop}%
\end{thebibliography}%

\end{document}